\documentclass[11pt,a4paper]{article}
\pdfoutput=1

\usepackage[compat=1.1.0]{tikz-feynman} 
\usepackage{dsfont}
\usepackage{jheppub}
\usepackage{rotating}
\usepackage{array}
\usepackage{amsmath}
\usepackage[normalem]{ulem}
\usepackage{slashed}
\usepackage{booktabs}
\usepackage{units}
\usepackage{multirow}
\usepackage{multicol}
\usepackage{url}
\usepackage{graphicx}
\usepackage{xspace}
\usepackage{subfig}
\usepackage{xcolor}
\usepackage{diagbox}
\usepackage{bbm}
\usepackage{lineno}
\usepackage{placeins}
\usepackage{mathtools}

\newcommand{\epem}{\ensuremath{e^+e^-}\xspace}

\newcommand{\beq}{\begin{equation}}
\newcommand{\eeq}{\end{equation}}

\newcommand{\commentOut}[1]{}
\newcommand{\commentout}[1]{}

\newcounter{savefootnote}

\def \belletwo {Belle\,II\xspace}

\def \superkekb {SuperKEKB\xspace}
\def \gazelle {GAZELLE\xspace}

\def \godzilla {GODZILLA\xspace}

\title{Physics reach of a long-lived particle detector at \belletwo}

\author[1]{S.~Dreyer,}

\author[1]{T.~Ferber,$^*$\footnote{\textit{now at: Karlsruhe Institute of Technology, 76131 Karlsruhe, Germany}}}
\emailAdd{torben.ferber@kit.edu}
\author[2]{A.~Filimonova,}
\author[1]{C.~Garcia-Cely,}
\author[3]{C.~Hearty,}
\author[1]{S.~Longo,}
\author[4]{R.~Sch\"afer,}
\author[1]{K.~Schmidt-Hoberg,}
\author[5]{M.~Tammaro,}
\author[6]{K.~Trabelsi,}
\author[4]{S.~Westhoff,$^*$}
\emailAdd{westhoff@thphys.uni-heidelberg.de}
\author[7]{J.~Zupan}

\affiliation[1]{Deutsches Elektronen--Synchrotron, 22607 Hamburg, Germany}
\affiliation[2]{NIKHEF, NL-1098 XG Amsterdam, The Netherlands}
\affiliation[3]{Department of Physics and Astronomy, University of British Columbia, Vancouver, British Columbia V6T 1Z1, Canada; Institute of Particle Physics (Canada), Victoria, British Columbia V8W 2Y2, Canada}
\affiliation[4]{Institute for Theoretical Physics, Heidelberg University, 69120 Heidelberg, Germany}
\affiliation[5]{Jo\v{z}ef Stefan Institute, Jamova 39, 1000 Ljubljana, Slovenia}
\affiliation[6]{Universit\'{e}  Paris-Saclay,  CNRS/IN2P3,  IJCLab,  91405  Orsay,  France}
\affiliation[7]{Department of Physics, University of Cincinnati, Cincinnati, Ohio 45221,USA}

\abstract{We have studied three realistic benchmark geometries for a new far detector \gazelle to search for long-lived particles at the \superkekb accelerator in Tsukuba, Japan. The new detector would be housed in the same building as \belletwo and observe the same \epem collisions. To assess the discovery reach of GAZELLE, we have investigated three new physics models that predict long-lived particles: heavy neutral leptons produced in tau lepton decays, axion-like particles produced in $B$ meson decays, and new scalars produced in association with a dark photon, as motivated by inelastic dark matter. We do not find significant gains in the new physics discovery reach of \gazelle compared to the \belletwo projections for the same final states. The main reasons are the practical limitations on the angular acceptance and size of \gazelle, effectively making it at most comparable to \belletwo, even though backgrounds in the far detector could be reduced to low rates. A far detector for long-lived particles would be well motivated in the case of a discovery by \belletwo, since decays inside \gazelle would facilitate studies of the decay products. Depending on the placement of \gazelle, searches for light long-lived particles produced in the forward direction or signals of a confining hidden force could also benefit from such a far detector. Our general findings could help guide the design of far detectors at future electron-positron colliders such as the ILC, FCC-ee or CEPC.}

\begin{document}

\maketitle

\section{Introduction}\label{sec:intro}
While strong theoretical arguments point towards the existence of physics beyond the Standard Model (SM), no unambiguous signal of new physics has been observed at high-energy colliders such as the Large Hadron Collider (LHC). There is no clear theoretical guidance for the scale of new physics, and many possibilities remain viable. 
In this situation, it is desirable to diversify the experimental search program as much as possible, taking full advantage of the current and future experimental facilities.

If new physics exists below the weak scale, strong constraints necessarily require rather weak couplings to the SM states. 
Light weakly coupled hidden sectors offer rich phenomenology of beyond-the-Standard Model signatures even in the minimal models, and have received significant attention over the last few years
~\cite{Batell:2009di,Batell:2009yf,Andreas:2012mt,Schmidt-Hoberg:2013hba,Essig:2013vha,Izaguirre:2013uxa,Morrissey:2014yma,Batell:2014mga,Dolan:2014ska,Krnjaic:2015mbs,Dolan:2017osp,Knapen:2017xzo,Beacham:2019nyx,Bernreuther:2019pfb,Bondarenko:2019vrb,Filimonova:2019tuy,BelleII:2020fag,Bernal:2017mqb,Baek:2020owl}. An intriguing possibility is that a light, feebly coupled dark sector contains the Dark Matter (DM) particle. This possibility is especially interesting in light of the stringent constraints on weak scale DM placed by direct detection experiments, which are already probing loop-suppressed scattering cross sections~\cite{Aprile:2018dbl,Ren:2018gyx}. Many of the light dark sector models naturally feature
long-lived particles (LLPs), due to small couplings and/or small mass splittings in the hidden sector.
The search for such states at colliders is complicated by the fact that for sufficiently long lifetimes the particles decay outside the active detector volume. In this limit only  generic signatures with missing energy remain.

To achieve a better coverage for such scenarios, adding a so-called far detector at a distance from the primary interaction point can enhance the sensitivity to long decay lengths compared to the main detector. At the
 LHC, this idea has motivated far detectors such as the recently approved 
 experiment FASER~\cite{Feng:2017uoz} or the proposed experiments CODEX-b~\cite{Gligorov:2017nwh}, MATHUSLA~\cite{Chou:2016lxi}, and ANUBIS~\cite{Bauer:2019vqk}. Far detectors at the LHC have been shown to provide a substantial gain in sensitivity for a large number of light and weakly coupled new physics scenarios.
 
At the \belletwo experiment at SuperKEKB in Japan, 
the conditions for LLP production and decay are different from the LHC case: LLPs are produced with a lower boost and in a cleaner environment with less background, making studies of LLPs with the \belletwo detector alone already very sensitive. Further improvements in the LLP signal reach could in principle be achieved using dedicated detectors.
The sensitivity to LLPs with very long decay lengths is mostly determined by the fiducial efficiency of a far detector, which scales with the angular acceptance times the radial thickness. Increasing the detector volume while keeping the full angular coverage would clearly enhance the sensitivity to LLPs beyond \belletwo. However, for practical reasons such as the positioning and supply of the \belletwo detector in the Tsukuba hall, it is impossible to build a far detector with full angular coverage.
To obtain realistic predictions for the physics reach of a far detector at \belletwo the instrumental requirements have to be investigated in a dedicated study.
 
\begin{figure*}[t!]
\begin{center}
\includegraphics[width=0.19\textwidth,angle=90]{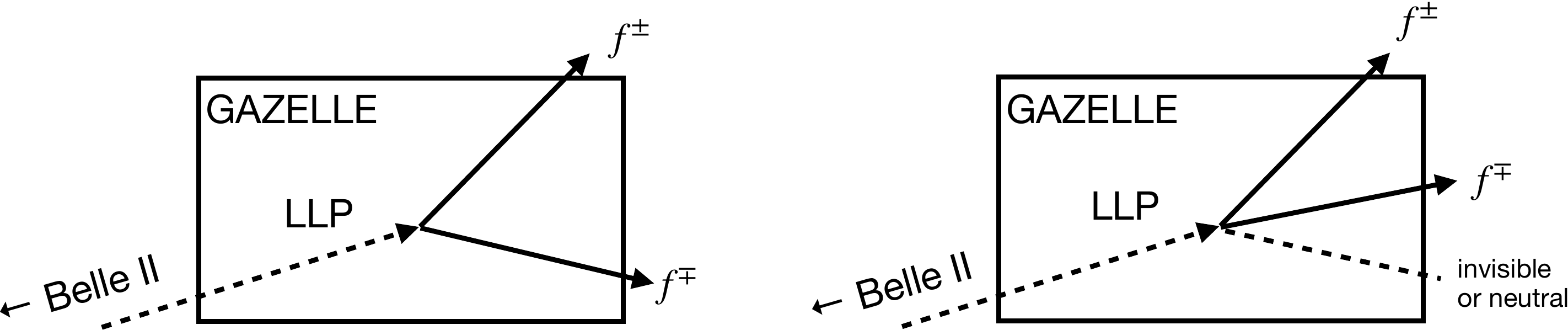}
\caption{Signature of signal events with two charged particles $(f^{\pm}=e^{\pm}, \mu^\pm, \pi^{\pm}, \mathrm{or\ } K^{\pm})$ (left) without and (right) with additional neutral particles or invisible particles.}
\label{fig:signal}
\end{center}
\end{figure*}

In this 
 paper, we explore the projected sensitivity of several realistic options for a far detector at \belletwo. We call the ensemble of new detectors \gazelle (\gazelle is the
Approximately Zero-background Experiment for Long-Lived Exotics).\footnote{GAZELLE is a recursive acronym. Recursive acronyms are common in computing and used by many organizations~\cite{recursive}.} For our study we focus on models with LLPs that decay into two charged particles, plus potential additional neutral particles (Fig.\,\ref{fig:signal}).
 The signature in \gazelle consists of two charged tracks that originate from the LLP decay vertex. Neutrinos and other neutrals escape the \gazelle detector as missing energy. Our goal is to assess the sensitivity gain of \gazelle compared to decays in the \belletwo tracking detectors for a series of benchmark models. In this way we can explore the discovery potential of a far detector at an $e^+ e^-$ experiment, which might also guide the construction of far detectors at future $e^+ e^-$ colliders like the ILC,  FCC-ee or CEPC.\\

The 
 paper is organized as follows. In Section~\ref{sec:detector} we describe the three \gazelle detector options and briefly comment on Belle\,II's tracking detectors, which will serve as a comparison. In Section~\ref{sec:backgrounds} we give a qualitative overview of possible backgrounds and discuss ways to control these backgrounds to achieve a good sensitivity to LLPs at \gazelle. Section~\ref{sec:models-intro} is devoted to dedicated sensitivity studies for three benchmark models: heavy neutral leptons\,(HNLs), axion-like particles\,(ALPs), and inelastic Dark Matter\,(iDM). We generalize our findings in Section~\ref{sec:generalaspects} to cover models with similar features. In Section~\ref{sec:exotica} we explore alternative new physics scenarios that could be detected with a \gazelle-like far detector at \belletwo. We finally conclude in Section~\ref{sec:summary} and give an outlook to far detectors at future $e^+e^-$ colliders.

\section{Detector}\label{sec:detector}
To measure the aforementioned LLP final states we require a detector that is capable of tracking charged particles and reconstructing a common vertex of those particles. For simplicity we restrict the discussion to a vertex within the air-filled \gazelle detector volumes. If this requirement is loosened, the fiducial volume of \gazelle can be increased at the expense of potentially larger backgrounds. \\

The physics requirements for the \gazelle detector are a timing resolution of the order of $100$\,ps to reconstruct the flight direction and velocity of particles, moderate latency times of the order of one microsecond to include \gazelle in the \belletwo trigger, and a position reconstruction precise enough to obtain a clear vertex resolution of the order of 10\,cm for an LLP that decays inside the \gazelle detector.  The previously defined physics requirements do not mandate calorimetry to measure photons, or very good position resolutions. This significantly reduces the cost of the \gazelle detector. Such a detector can provide the following observables that can be used to reconstruct the LLP kinematics and to reject backgrounds:
\begin{itemize}
    \item Vertex: Two charged tracks enable the reconstruction of a common decay vertex. This allows a very precise direction measurement of the LLP direction given the long distance of \gazelle from the \epem interaction point.
    \item Mass: The LLP mass can be reconstructed from the track pair opening angle and the speed $\beta$ of each track. $\beta$ is measured using the precise timing information along the track \cite{Curtin:2017izq}.
    \item Track direction: The sign of $\beta$ (or the
    direction of the track) can be used to reject backgrounds from tracks entering \gazelle.
    \item Pointing angle: Since LLPs are moderately boosted, the two charged tracks can be used to reconstruct an approximate pointing angle that must be consistent with an LLP origin near the \epem collision point.
    \item Absolute time: When synchronized with the \belletwo readout, a loose time coincidence of activity in both detectors can be used to reject backgrounds.
\end{itemize}
Since it will be challenging to provide a magnetic field for such large volumes, it is not possible to directly reconstruct the particle momentum in \gazelle.\\

\begin{figure}[ht]
\begin{center}
\includegraphics[height=15cm, angle=90]{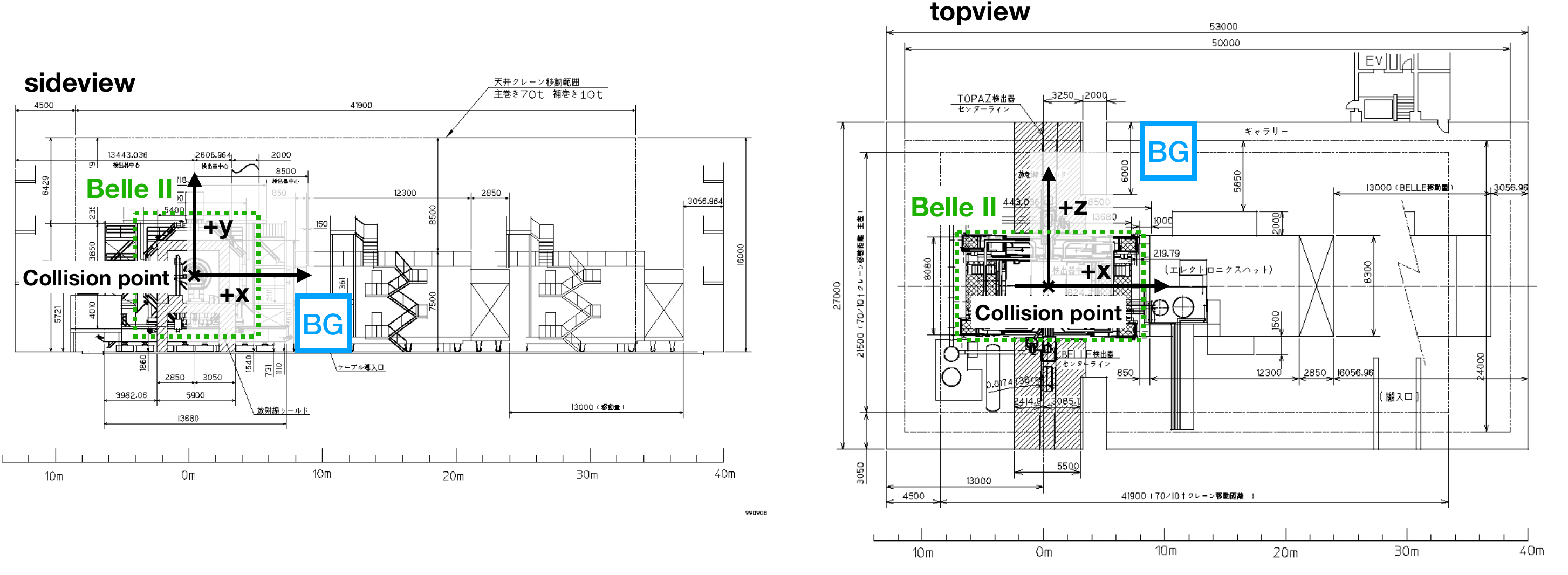}
\caption{
Baby-\gazelle (left) sideview and (right) topview. 
}\label{fig:det:baby} 
\end{center}
\end{figure}

\begin{figure}[ht]
\begin{center}
\includegraphics[height=15cm, angle=90]{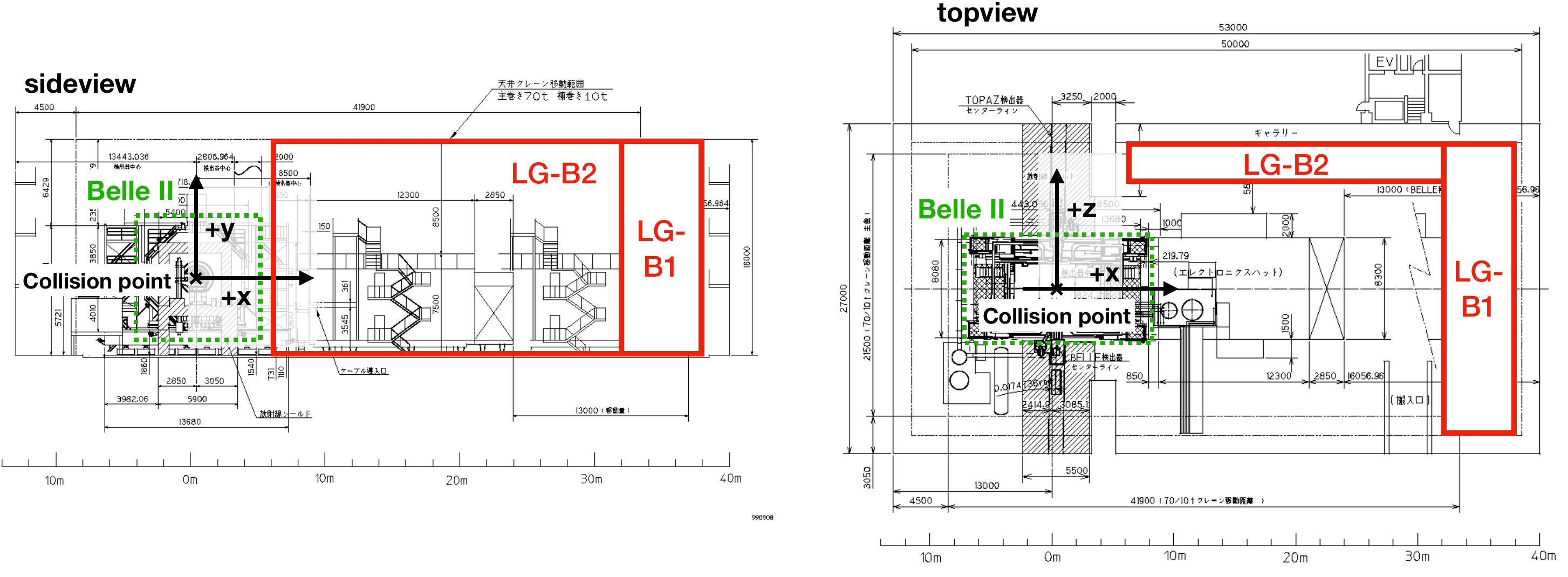}
\caption{
L-\gazelle  (left) sideview and (right) topview. 
}\label{fig:det:lgz} 
\end{center}
\end{figure}

\begin{figure}[ht]
\begin{center}
\includegraphics[height=12cm, angle=90]{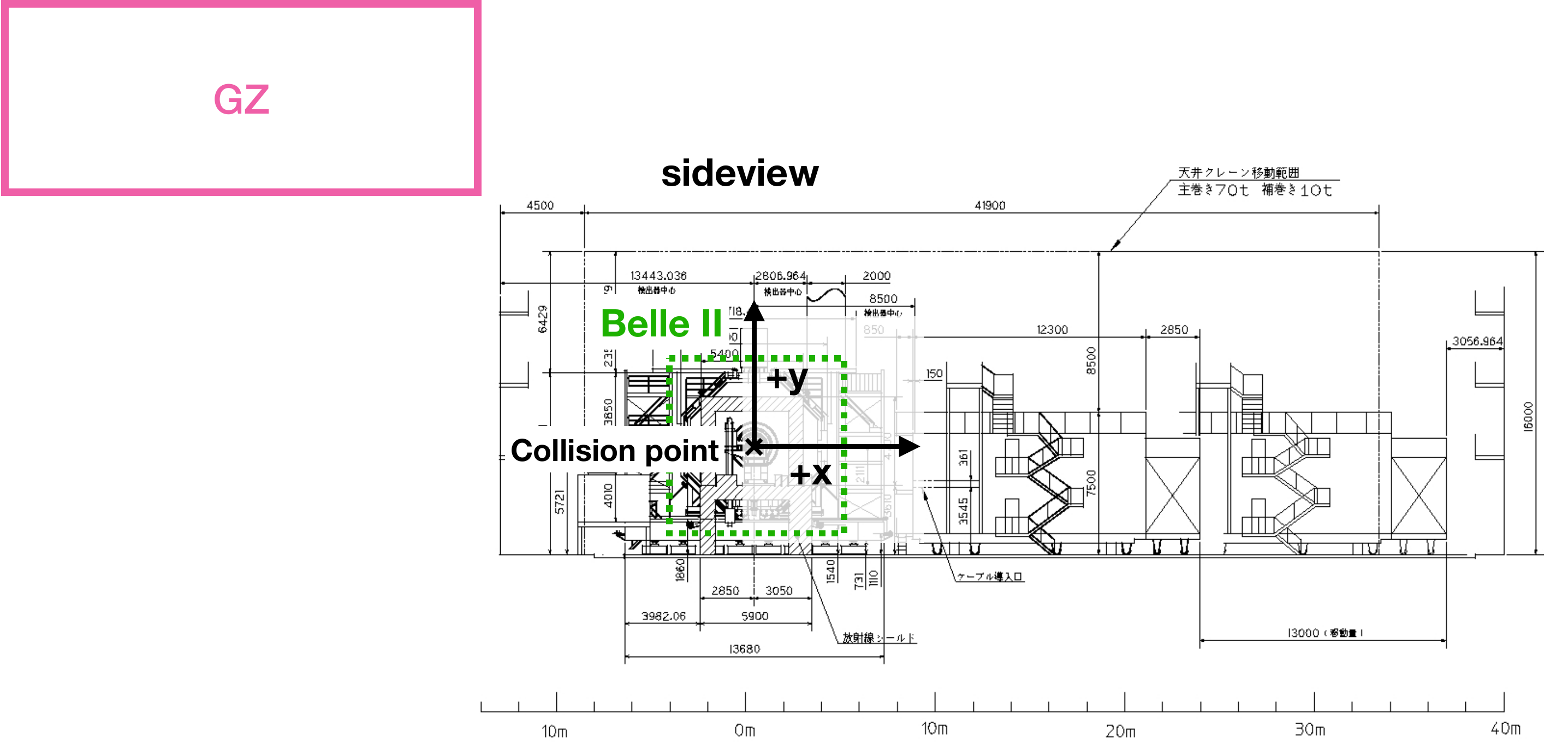} \\[1cm]
\includegraphics[height=12cm, angle=90]{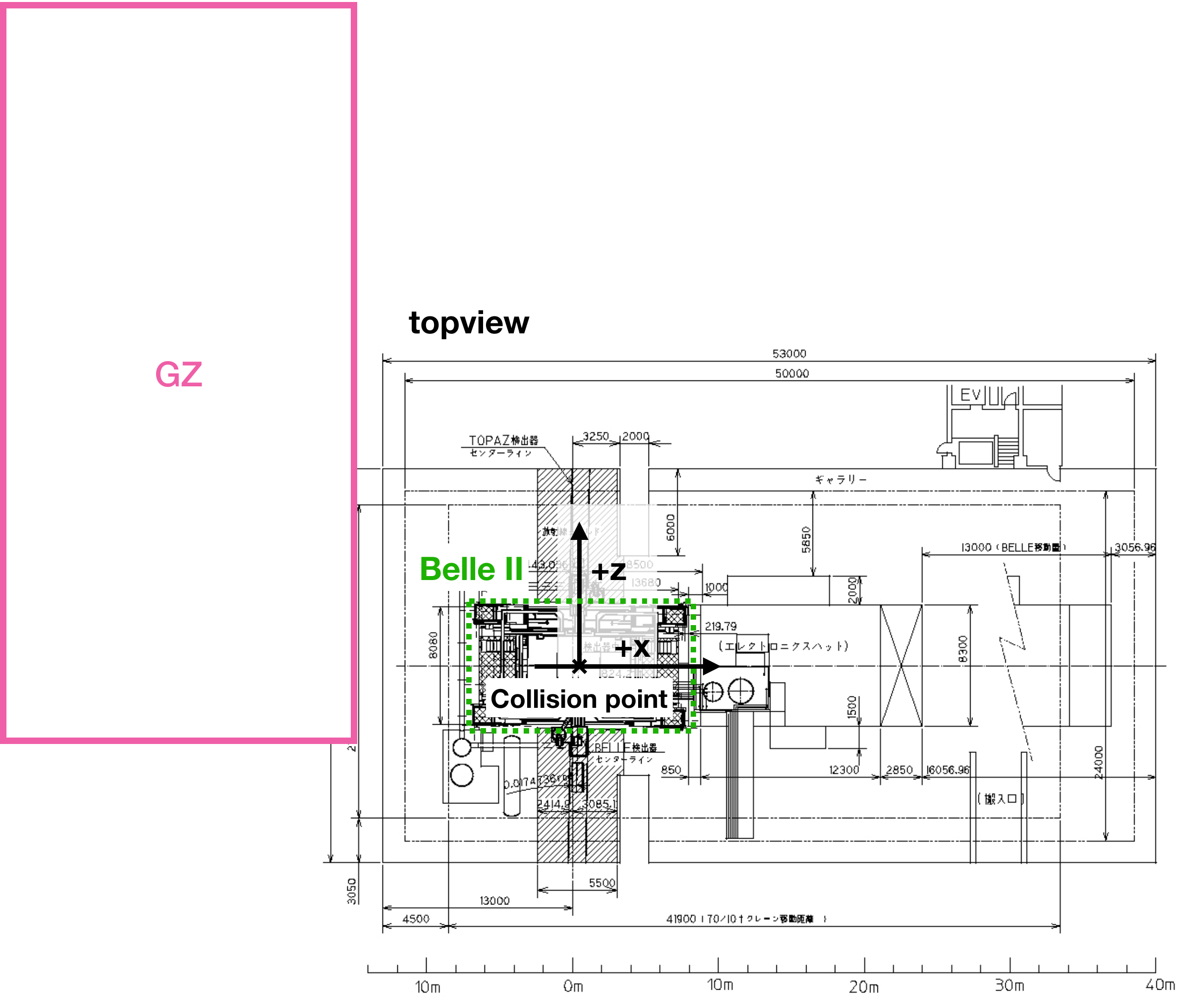}
\caption{
GODZILLA (top) sideview and (bottom) topview.
}\label{fig:det:godzilla} 
\end{center}
\end{figure}

We focus on four detector configurations that differ in size and positioning.  The coordinate system is oriented such that the \belletwo collision point is at $(0\,\text{m}, 0\,\text{m}, 0\,\text{m})$. The $z$--axis of the laboratory frame coincides with that of the \belletwo solenoid and its positive direction is approximately that of the incoming high--energy electron beam. The \gazelle detector configurations are:
\begin{itemize}
    \item Baby-\gazelle (BG): A $(4\times4\times4)\,\text{m}^3$ cube-shaped detector positioned on the floor of Tsukuba hall centered at $x \approx $\,10\,m, $y \approx $-3.7\,m, $z \approx $\,10\,m (Fig.\,\ref{fig:det:baby}). The solid angle coverage is $\Omega = 0.12$\,sr (0.95\,\% angular coverage) 
    \item L-\gazelle (LG, consisting of LG-B1 and LG-B2): 
    \begin{itemize}
        \item LG-B1: A $(6\times16\times24)\,\text{m}^3$ detector covering the forward wall of Tsukuba hall centered at $x \approx 35$\,m, $y \approx 2.3$\,m, $z \approx 0$\,m (Fig.\,\ref{fig:det:lgz}). The solid angle coverage is  $\Omega = 0.34$\,sr (2.7\,\% angular coverage) 
        \item LG-B2: A $(26\times16\times3)\,\text{m}^3$ detector covering the far wall of Tsukuba hall centered at $x \approx 19$\,m, $y \approx 2.3$\,m, $z \approx 10.5$\,m (Fig.\,\ref{fig:det:lgz}). The solid angle coverage is $\Omega = 0.76$\,sr (6.0\,\% angular coverage) 
    \end{itemize}
    \item GODZILLA (GZ): A $(25\times10\times50)\,\text{m}^3$ detector placed outside of Tsukuba hall at ground level centered at $x \approx -27$\,m, $y \approx 18$\,m, $z \approx 20$\,m (Fig.\,\ref{fig:det:godzilla}). The solid angle coverage is $\Omega = 0.74$\,sr (5.9\,\% angular coverage) 
\end{itemize}

The placements could be realized with minimal or moderate civil engineering around the \belletwo interaction point. The exact placement would be subject to civil engineering constraints such as existing doors or the \belletwo electronics hut.\\

There are several  options for a detector technology that fulfills the aforementioned requirements, while at the same time remaining cost efficient. These include detectors based on scintillating fibers, scintillating plates, and multi-gap resistive plate chambers (RPCs):
\begin{itemize}
    \item Scintillating fibers are a well tested option that are being used for the SciFi tracker upgrade of the LHCb detector  \cite{Hopchev:2017tee}. One could adopt this layout, where the detector is built from individual modules ($0.5\,$m $\times$ $4.8\,$m), each comprising 8\,fibre mats with a length of $2.4\,$m as active detector material. The fibre mats consist of 6\,layers of densely packed blue-emitting scintillating fibres with a diameter of 250$\,\mu$m. The scintillation light could be read out by multi-channel photomultipliers or SiPMs. 
     \item Another option would be low-cost scintillating plates, such as the scintillating tiles used as an active medium in the TileCal hadronic calorimeter of the ATLAS detector \cite{CERN-LHCC-2017-019}, or the long organic plastic scintillator strips that are widely used in many neutrino experiments (e.g. MINOS, OPERA, MINERvA, T2K).
      \item Multigap Resistive Plate Chambers (MRPC) represent another fairly inexpensive option with good timing and spatial resolution.  This technology has been employed in the ALICE \cite{Fonte:1999cb} and STAR \cite{Yang:2014xta} detectors. 
\end{itemize}
For the studies in this paper we do not assume a particular detector technology but assume full detection efficiency within the respective \gazelle volumes for charged particles.\\

The \gazelle detector should be synchronized with the \belletwo readout and ideally even with the level 1 trigger system in both directions, i.e. triggering \belletwo with \gazelle and vice-versa. While offline synchronisation of the two detectors can be used to reject background and reconstruct the LLP production mechanisms, low multiplicity final states such as those discussed in Sec.\,\ref{sec:IDM}, might not be triggered by the default \belletwo trigger algorithms.\\

We compare the sensitivity of \gazelle to the sensitivity that can be obtained from LLP decays within the \belletwo tracking detectors. We assume that the \belletwo tracking detectors are fully efficient for polar angles $17^{\circ} < \theta < 150^{\circ}$, $-55\,\rm cm < z < 140\,\rm cm$ ($z$~as defined above), and radial displacements of up to 60\,cm. The solid angle coverage is $\Omega=11.5\,\text{sr}$ (90 \% angular coverage).

To reject backgrounds from prompt standard model decays, we exclude vertices with radial displacements up to 0.9\,cm. We use this selection whenever we show \belletwo results in the following sections.

\section{Backgrounds}\label{sec:backgrounds}
Any signature that can fake a two-track signature pointing back to the \belletwo interaction point and a timing consistent with a bunch crossing after correcting for time-of-flight is a potential background. There are two main classes of backgrounds in \gazelle: Those induced by the primary \epem collisions, and those from cosmic rays. The most abundant backgrounds are expected to be neutral $K^0_L$ mesons that decay into two charged particles and additional (undetected) neutral particles, and muon decays into one electron and two neutrinos.\\

\begin{figure*}[h]
\begin{center}
\includegraphics[width=0.4\textwidth,angle=90]{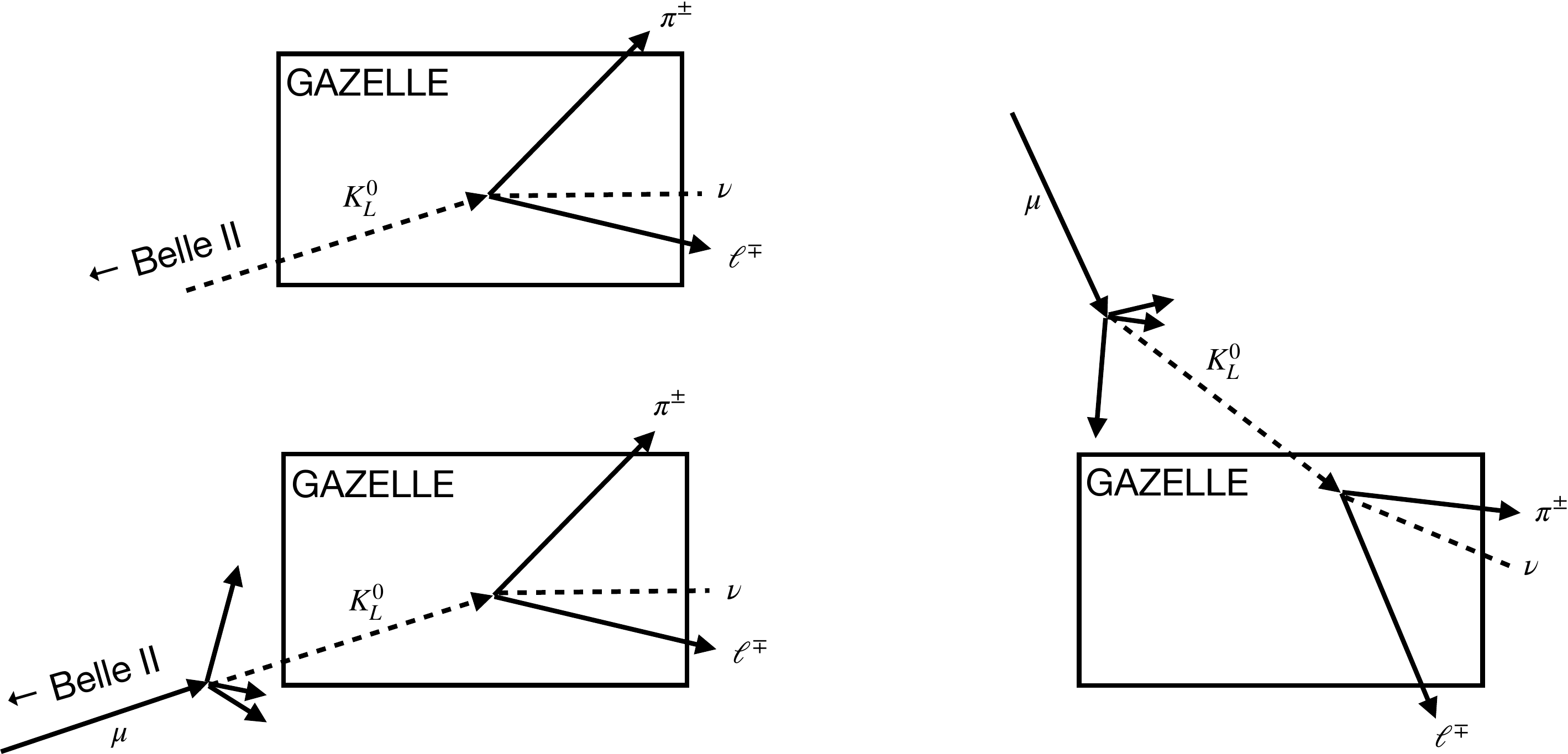}
\caption{Background from direct and muon-induced $K^0_L$ meson decays originating from (left) primary collisions and (right) cosmic muons.}
\label{fig:background_signature_k0l}
\end{center}
\end{figure*}

$K^0_L$ mesons originating from the \belletwo interaction point are almost completely absorbed in \belletwo or the surrounding concrete shielding. 
However, $K^0_L$ mesons can also be produced by muons originating directly from the interaction point or by muons from $B$, $D$, or $\tau$ decays, interacting with the shielding material. 
This background can be reduced if the primary muon is tracked in \belletwo (Fig.\,\ref{fig:background_signature_k0l} left). To a lesser extent muon-induced production of $K^0_S$ mesons and $\Lambda$ baryons will produce similar backgrounds if they are produced close to the \gazelle detector. While neutral hadrons are a rather rare component in cosmic rays, the same muon-induced production process can occur in the material surrounding the  Tsukuba hall. The incoming cosmic muon cannot be used to reject this background since it never enters any active detector, but the pointing angle of the $K^0_L$ meson  decay products will generally not point towards \belletwo (Fig.\,\ref{fig:background_signature_k0l} right).

\begin{figure*}[h]
\begin{center}
\includegraphics[width=0.225\textwidth,angle=90]{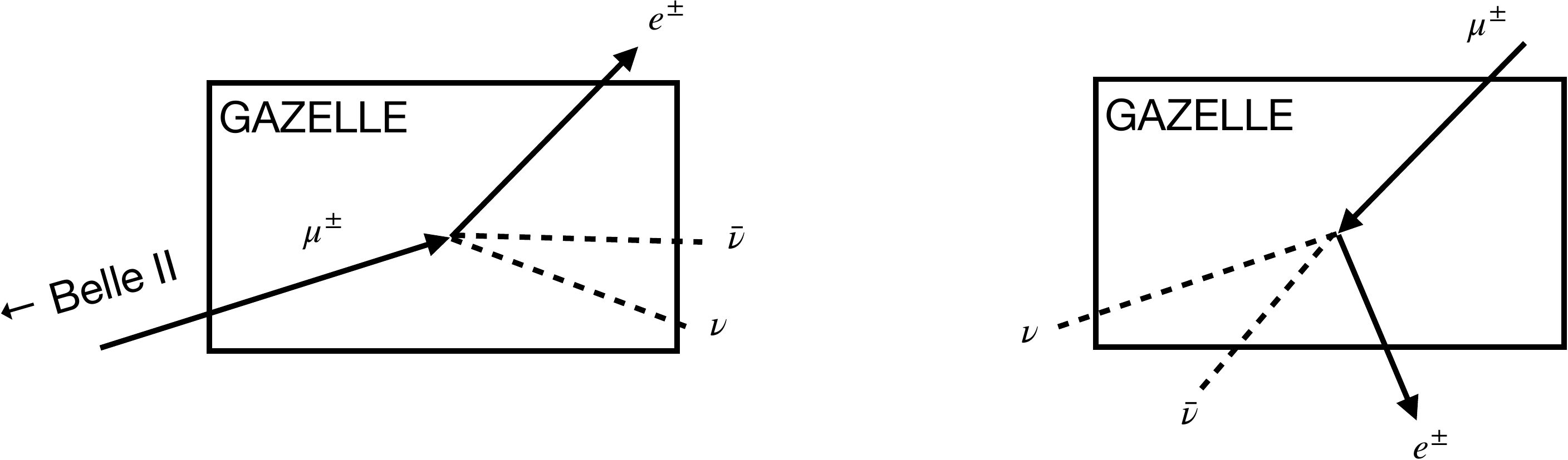}
\caption{Background from (left) muon decays from primary collisions and (right) cosmic muons.}
\label{fig:background_signature_muondecays}
\end{center}
\end{figure*}

Muon decays can mimic a 2-body particle decay if the incoming muon direction is not correctly reconstructed (Fig.\,\ref{fig:background_signature_muondecays}). In that case a muon decay would appear as a LLP decay into two charged particles. While for muons from \belletwo the decay kinematics typically points away from \belletwo and can be rejected, cosmic muons can mimic LLP decays that point towards \belletwo .

There are several features of \gazelle that would allow for potent rejection of these backgrounds.
\begin{itemize}
    \item The direction and angles of the particles in relation to the Belle II interaction point will typically result in a reconstructed mass that is too large to have been produced in \superkekb collisions.
    \item A two-body LLP decay into two charged particles would ensure that the two charged particle tracks and the line connecting the decay vertex to the collision point lie in a plane. This kinematic feature would be extremely rare among cosmic muon induced events. 
    \item  LLPs are typically produced together with other prompt particles that are visible in \belletwo, for example from $B$ or $\tau$ decays. If the event missing momentum vector can be computed in \belletwo, it must point to the vertex position in \gazelle. 
    \item The event in \gazelle and \belletwo must have a time coincidence. Since the LLP mass reconstruction is model-dependent, this timing window must be rather large but it will still allow powerful background rejection.
    
\end{itemize}

While additional passive shielding between \belletwo and \gazelle stops neutral hadrons, it also increases the rate of muon-induced hadron production and must be optimized accordingly. If the shielding is too close to the \gazelle volume, additional background from $K^0_S$ meson decays into two pions, as well as $\Lambda$ decays, are a concern. Active cosmic vetoes on top of \gazelle could be used to further reject cosmic muon induced backgrounds.\\

Preliminary simulations of the \belletwo hall and simplified detector setups show that cosmic muon decays and meson decays from \epem collisions are a particular experimental challenge. 
A full study of the backgrounds is beyond the scope of this paper. 
In the following, we assume an optimistic scenario with zero backgrounds, expecting that the combination of background rejection measures discussed above will be able to reduce the GAZELLE backgrounds to a reasonable level.

\section{Benchmark models with long-lived particles}\label{sec:models-intro}
To assess the potential discovery reach of the different \gazelle detectors, we focus on three simple models: heavy neutral leptons (HNLs) with couplings to $\tau$ leptons; axion-like particles (ALPs); and dark scalars in a model for inelastic dark matter(iDM). These three models are distinguished by the production of an LLP. HNLs are produced in $\tau$ lepton decays, ALPs are produced in meson decays, and dark scalars from inelastic dark matter are produced in association with a dark photon, similar to Higgs-$Z$ associated production at a future electron-positron collider. While other production modes are possible in each model, we focus on these three modes, which lead to different kinematics of the LLP decay products and resemble most of the common ways to produce LLPs at \belletwo. We consider the three selected models as representatives of classes of models with similar features.

The three models exploit the main assets of \belletwo: producing tau leptons in a clean environment (HNLs); probing extremely rare $B$ meson decays (ALPs); and resolving displaced vertices (iDM). While other experiments share some of these features, we expect that Belle II is competitive or superior in its sensitivity to the corresponding signatures. Our goal is to explore to what extent GAZELLE could enhance the reach and thus strengthen the position of \belletwo in probing these and similar models.

In the following, we will discuss the discovery potential of \gazelle for each of the three models and compare it to the expected reach of \belletwo. In each model, the signature consists of two displaced charged particles from a common vertex, the decay products of the LLP.
 The detection of additional particles in the final state can enhance the sensitivity, but renders the search more model-dependent. Our goal is not to optimize the discovery potential for a single model, but to cover a wide class of models with similar signatures.

The long-lived ALPs and dark scalars decay into displaced pairs of leptons or into hadrons if these decay modes are kinematically accessible. Typically all decay products are detectable, and their invariant mass reconstructs to the mass of the decaying LLP. HNL decays always involve a neutrino in the final state. The missing energy prevents a direct reconstruction of the HNL's mass from the visible decay products. As a consequence, the reconstruction and background suppression is more involved than in the other two models.

To determine the probability that an event is seen in the \belletwo or GAZELLE detectors, we generate sample-sets of $N$ events using \texttt{EvtGen}\,\cite{Lange:2001uf} (ALPs) or \texttt{MadGraph}\,\cite{Alwall:2014hca} (HNLs and iDM). For each event $i$ in one of these samples, we extract the production point and the momentum direction of the LLP. If the LLP misses the detector as a whole, there is no chance to detect its decay products. Therefore we first check whether the LLP's direction geometrically intersects with the detector. If that is the case, we determine the distances $\ell^{in}$ and $\ell^{out}$ between the LLP's production point and the point where the LLP enters and leaves the detector. The probability $\mathds{P}$ for the LLP in event $i$ to decay inside the detector depends on its boost $\gamma\beta_i$ and proper decay length $c\tau$ as
\begin{align}\label{eq:LLP:probability}
    \mathds{P}_i= \exp\left(-\frac{\ell_i^{in}}{\gamma\beta_i c\tau}\right) - \exp\left(-\frac{\ell_i^{out}}{\gamma\beta_i c\tau}\right).
\end{align}
To obtain the average probability $\langle\mathds{P}\rangle$ for an LLP to decay inside the detector, we take the mean of all probabilities $\mathds{P}_i$ in the sample-set,
\begin{align}\label{eq:averageprob}
    \langle\mathds{P}\rangle &= \frac{1}{N}\sum_{i=1}^N \mathds{P}_i\,.
\end{align}
We calculate this probability for each of the three \gazelle detectors and for \belletwo, as described in Section~\ref{sec:detector}, for a few benchmark parameters in each LLP model.  Throughout our analysis we assume a detection efficiency of 100\% and zero background in the detector.

\subsection{Heavy neutral leptons}\label{sec:HNL}
\noindent Heavy neutral leptons (HNLs) are electroweak singlet fermions, $N$, of mass $m_{N}$ \cite{DREWES_2013}. While several of these new particles could exist, we restrict ourselves to the case of a single HNL. The HNL couples to the SM lepton doublets, $L_\alpha$, via the neutrino portal interaction
\beq
{\cal L}_{\rm HNL}=\sum_{\alpha} c_{\alpha} (\bar L_\alpha \tilde H) N',
\eeq
where the index $\alpha=1,2,3$ runs over the SM generations, $H$ is the Higgs doublet and $c_\alpha$ is a dimensionless coupling constant. This structure fixes the SM charges of $N'$ to be zero, thus $N'$ is also called a ``sterile neutrino''. After electroweak symmetry breaking the neutrino portal interaction generates a mixing between the SM neutrinos, $\nu_\alpha$, and the HNL, $N$. The neutrino flavor eigenstates, $\nu_\alpha'$, $N'$, and the mass eigenstates $\nu_\alpha, N$, are related by
\begin{align}
\nu_\alpha' \simeq \nu_\alpha+ U_\alpha N,\qquad
N' \simeq - U_\alpha \nu_\alpha + N,
\end{align}
to leading order in the small mixing angles $U_\alpha$.
This means that the interactions of $N$ are exactly the same as for the SM neutrinos,
rescaled by the mixing $U_{\alpha}$. If $N$ predominantly mixes with just a single SM lepton flavor, $\alpha$, we denote it as $N_\alpha$, \emph{i.e.}, for $N_e$ we have $U_e\gg U_{\mu,\tau}$.

\begin{figure}[t!]
	\centering
	\begin{tikzpicture}
	\begin{feynman}
	\vertex (tau) {\(\tau^-\)};
	\vertex[right=1.8cm of tau, dot] (vertex) {};
	\vertex[right=3cm of tau] (a);
	\vertex[above=3em of a] (nu) {\(N\)};
	\vertex[below=2em of a] (vertex2);
	\vertex[right=1.5cm of vertex2] (lp) {\(\bar \nu_\ell, \bar u\)};
	\vertex[below=of vertex2] (lm)  {\(\ell^-, d\)};
	
	\diagram* {
		{
			(tau) -- (vertex) -- (nu),
			(vertex2) -- (lp),
			(vertex2) -- (lm),
		},
		(vertex) -- [boson, edge label=\(W^-\)] (vertex2),
	};	
	\end{feynman}
	\end{tikzpicture}\hspace{1cm}
	\begin{tikzpicture}
		\begin{feynman}
		\vertex (tau) {\(N\)};
		\vertex[right=1.8cm of tau, dot] (vertex) {};
		\vertex[right=3cm of tau] (a);
		\vertex[above=3em of a] (nu) {\(\nu_\tau\)};
		\vertex[below=2em of a] (vertex2);
		\vertex[right=1.5cm of vertex2] (lp) {\(\bar f\)};
		\vertex[below=of vertex2] (lm)  {\(f\)};
		
		\diagram* {
			{
				(tau) -- (vertex) -- (nu),
				(vertex2) -- (lp),
				(vertex2) -- (lm),
			},
			(vertex) -- [boson, edge label=\(Z\)] (vertex2),
		};	
	\end{feynman}
	\end{tikzpicture}
	\caption{Representative Feynman diagrams for (left) tau-flavored HNL production via tau decay and (right) HNL decay. The dot vertex indicates the neutrino mixing parametrized by an angle $U_\tau$.}\label{fig:HNLdiagrams}
\end{figure}
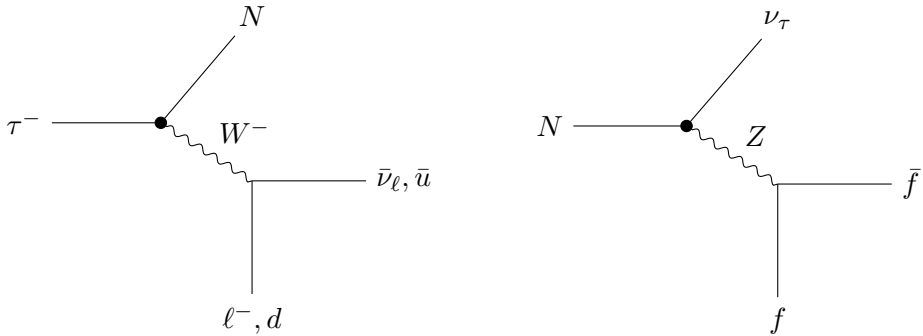


At \belletwo, several production mechanisms for HNLs can be important.
HNLs lighter than the $\tau$ lepton can be produced in $\tau$ decays (Fig.~\ref{fig:HNLdiagrams} left), due to the large number of $\tau$'s produced at \belletwo.
For heavier HNLs, the decays $B^\pm\to N_\alpha \ell^\pm$ and $B\to N_\alpha D \ell^\pm$ are the most important production processes. 
%
 We restrict our analysis to the case of $\tau$-flavored HNLs $N_\tau$, \emph{i.e.}, we take $U_\tau\ne 0$ and $U_e=U_\mu=0$, and assume that $m_N < m_\tau$.  In this region of parameter space the main source of HNLs are the decays $\tau\to N_\tau \ell \nu, N_\tau q\bar q$; other production channels can be safely neglected. We focus on the hadronic 2- and 3-body decays $\tau^\pm\to N_\tau\pi^\pm(\pi^0)$, since these contain easily identifiable final-state particles.  

Inside the \belletwo detector the HNL can decay via an off-shell $W$ or $Z$ boson, giving rise to the channels $N_\tau\to \nu \bar \nu \nu, \ell^+\ell^- \nu, \mathrm{or\ } q \bar q \nu$, with $\ell = e$ or $\mu$ (Fig.~\ref{fig:HNLdiagrams} right). We consider all kinematically available decay channels for a given $m_N$ and require at least two charged particles in the final state. 
Analytical expressions for $N$ production and decay are taken from Ref.~\cite{Bondarenko:2018ptm} and are summarized in Appendix~\ref{app:HNL}.

\begin{figure}[t]
\begin{center}
\includegraphics[width=.55\linewidth]{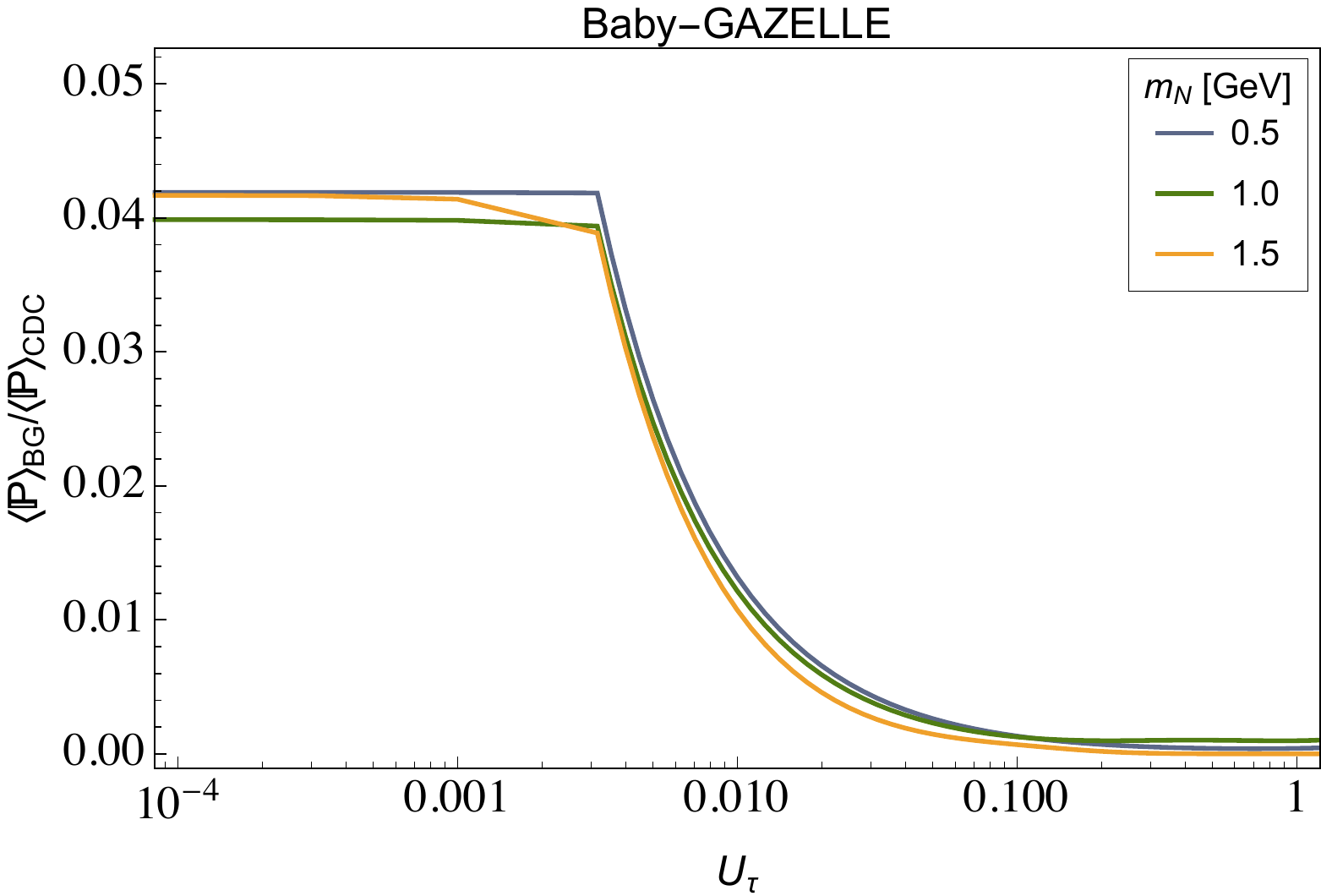}\\[1cm]
\includegraphics[width=.55\linewidth]{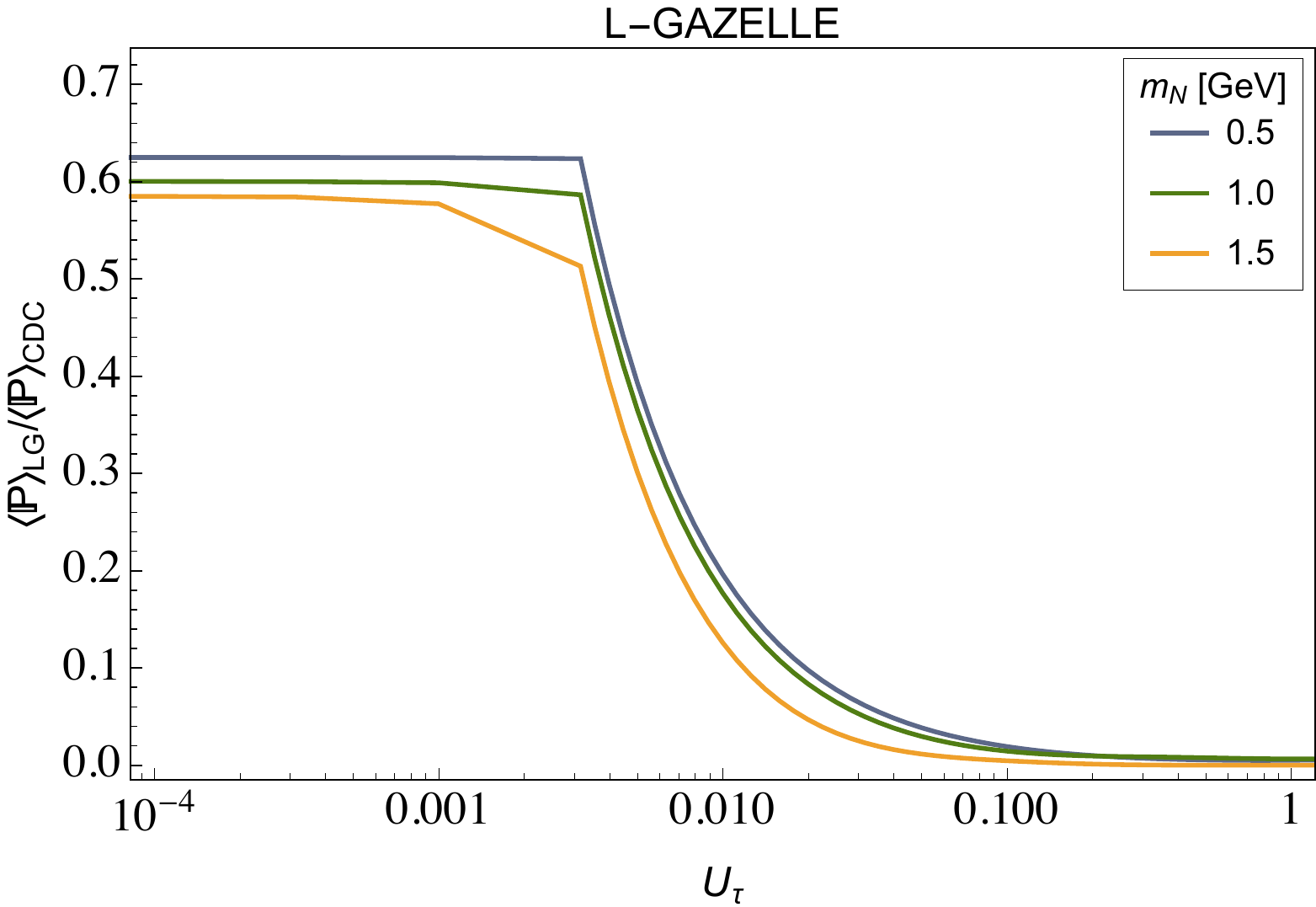}\\[1cm]
\includegraphics[width=.55\linewidth]{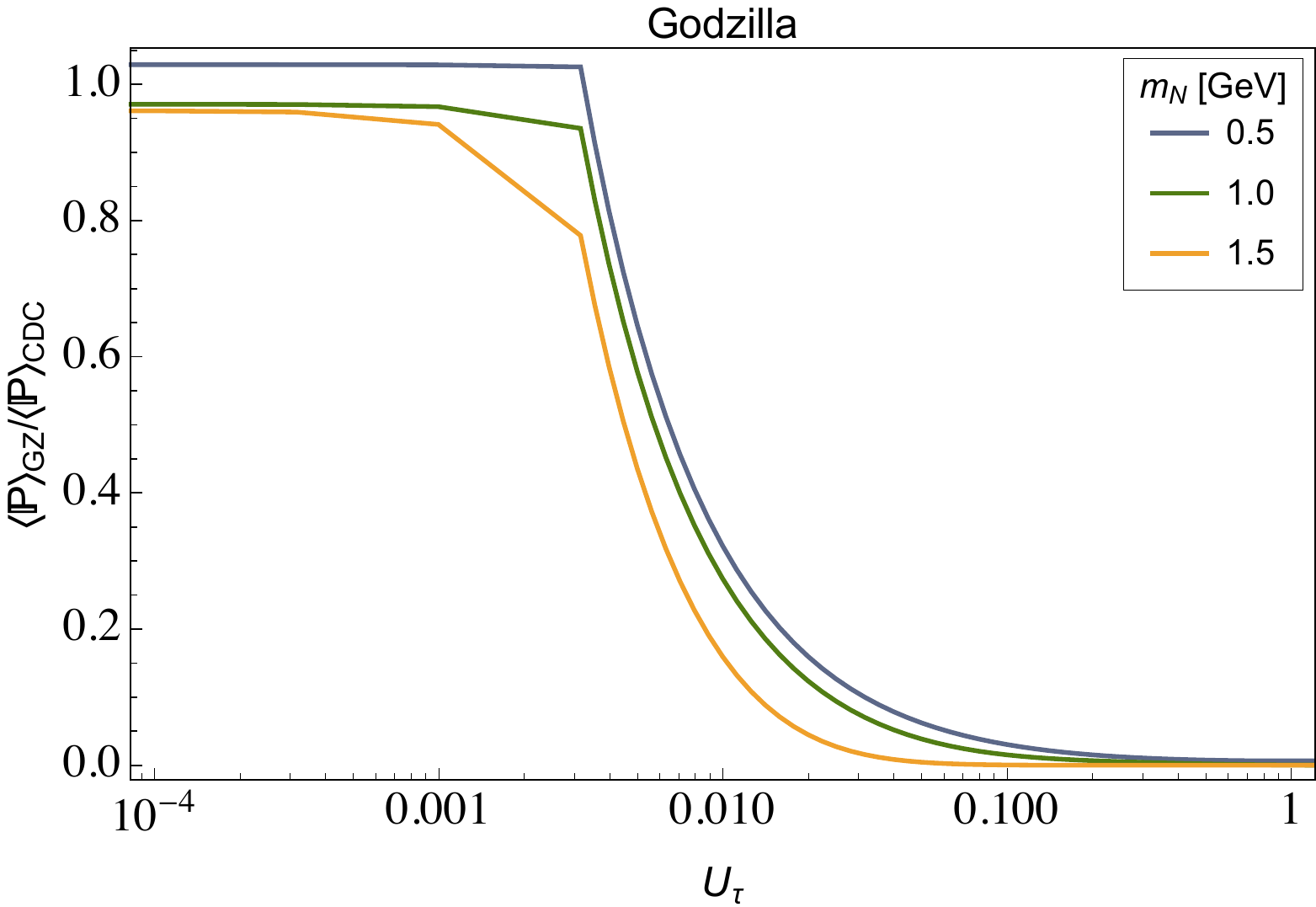}
\caption{Ratio of average decay probabilities for a HNL 
 $N_\tau$ within different configurations of \gazelle/\belletwo, as a function of the mixing angle $U_\tau$, for three HNL benchmarks.}\label{fig:prob:HNL}
\end{center}
\end{figure}

We have simulated events at \belletwo using \texttt{MadGraph} for three benchmark HNL masses, $m_N = 0.5,\,1.0,\,1.5$ GeV. Using the full kinematical information, we compute the average probability of an HNL to decay inside the detector volume as in Eq.~\eqref{eq:LLP:probability}. Fig.~\ref{fig:prob:HNL} shows the average decay probability for each GAZELLE configuration, normalized to the average probability to decay within \belletwo.

We then estimate the expected number of HNL decays inside the \gazelle detector as
\beq
N_{\rm dec} = N_{\tau\tau} \times{\rm Br}_{\rm prod}(m_N,U_\tau) \times{\rm Br}_{\rm dec}(m_N) \times \langle\mathds{P}(m_N,U_\tau)\rangle\,.
\eeq
Here $N_{\tau\tau} = 5\times10^{10}$ is the number of $\tau$ pairs produced at \belletwo with $50$ ab$^{-1}$ luminosity. The branching ratio for the inclusive decay $\tau\to N X$, ${\rm Br}_{\rm prod}(m_N,U_\tau)$, depends on both the HNL mass and the mixing angle with the tau neutrino. Further, ${\rm Br}_{\rm dec}(m_N)$ is the branching ratio for the inclusive decay $N\to \nu X_{\rm dec}$, where $X_{\rm dec}$ contains at least two visible particles, \textit{i.e.}, charged leptons or pions.

\begin{figure}[t]
\begin{center}
\includegraphics[width=.575\linewidth]{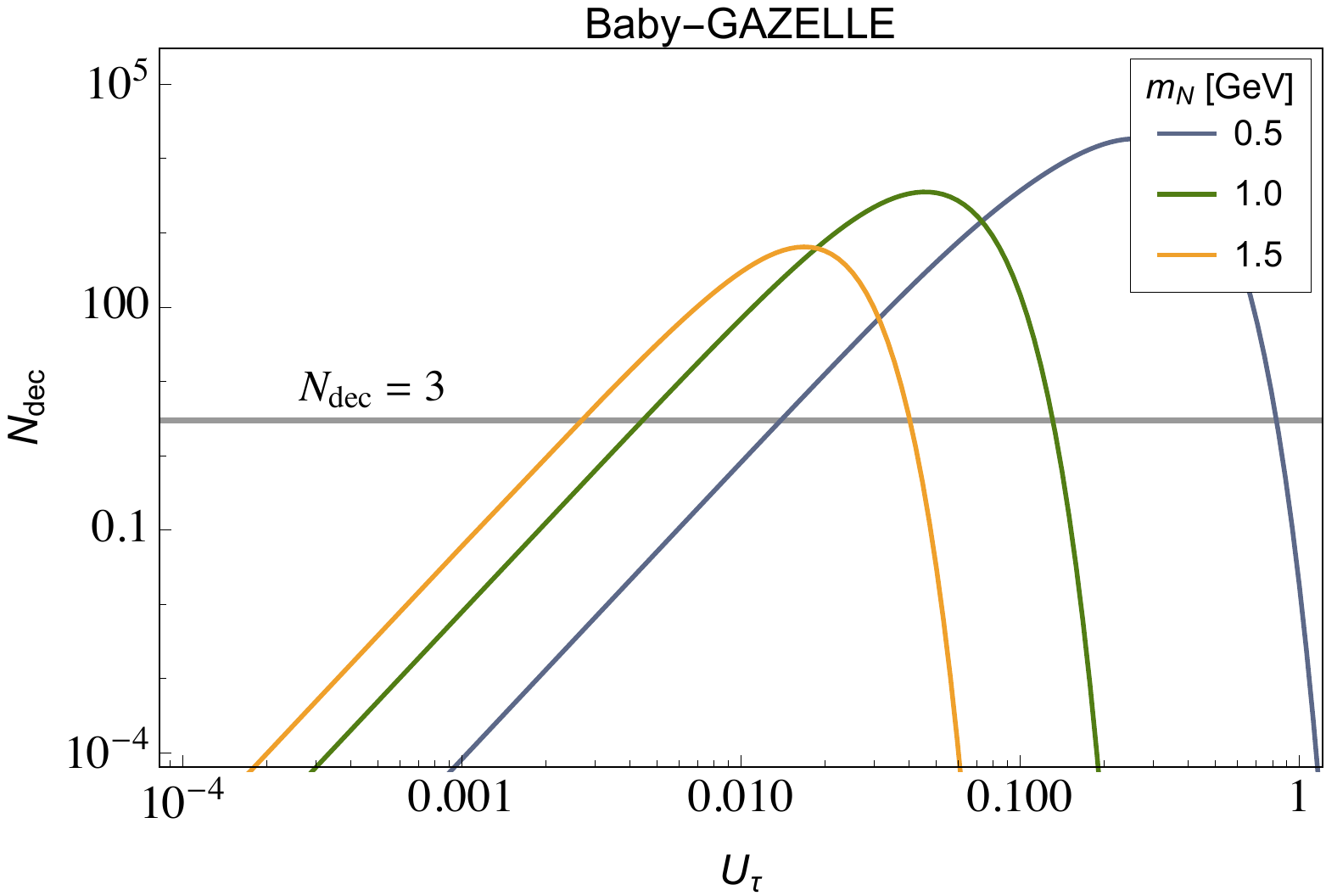}\\[1cm]
\includegraphics[width=.575\linewidth]{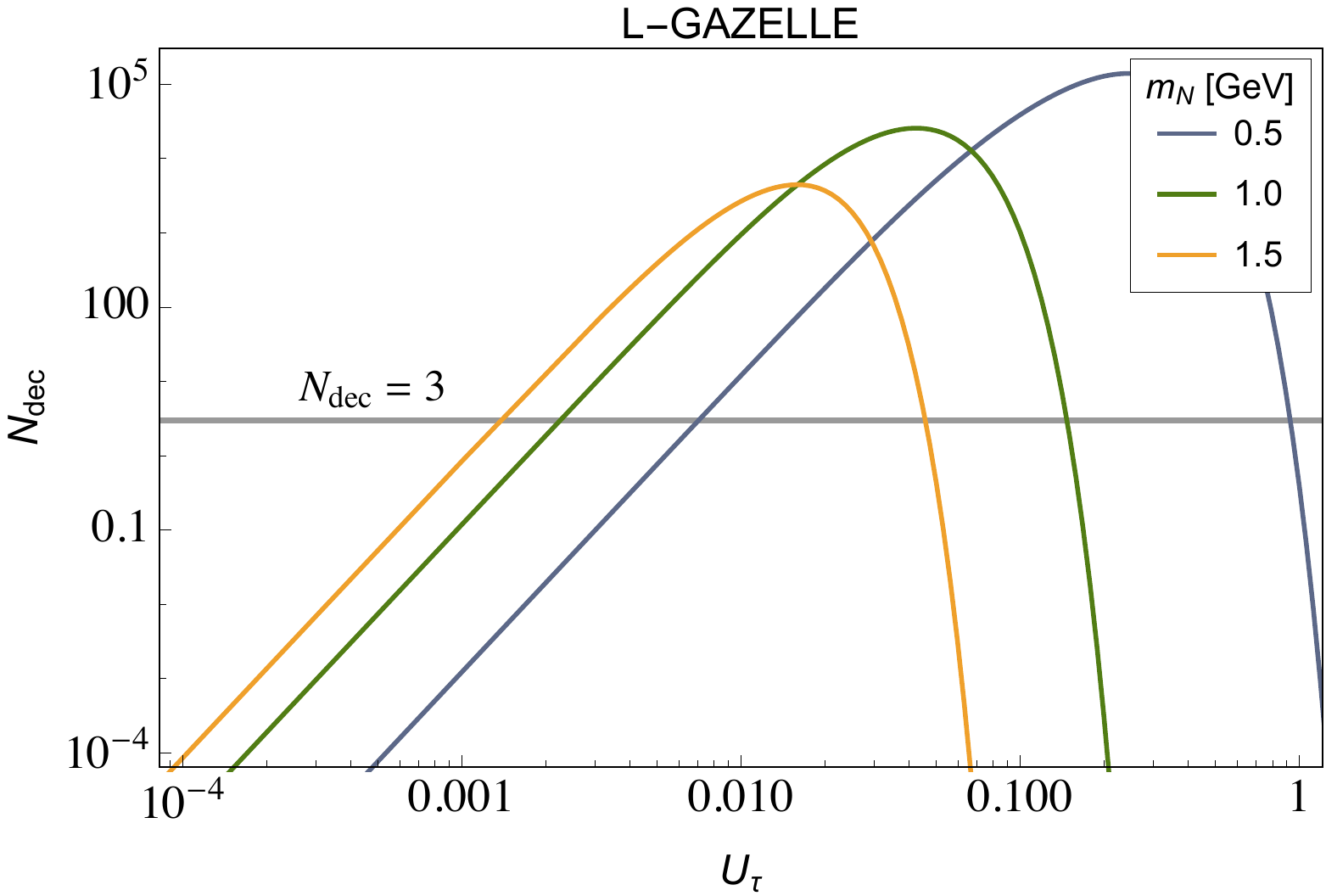}\\[1cm]
\includegraphics[width=.575\linewidth]{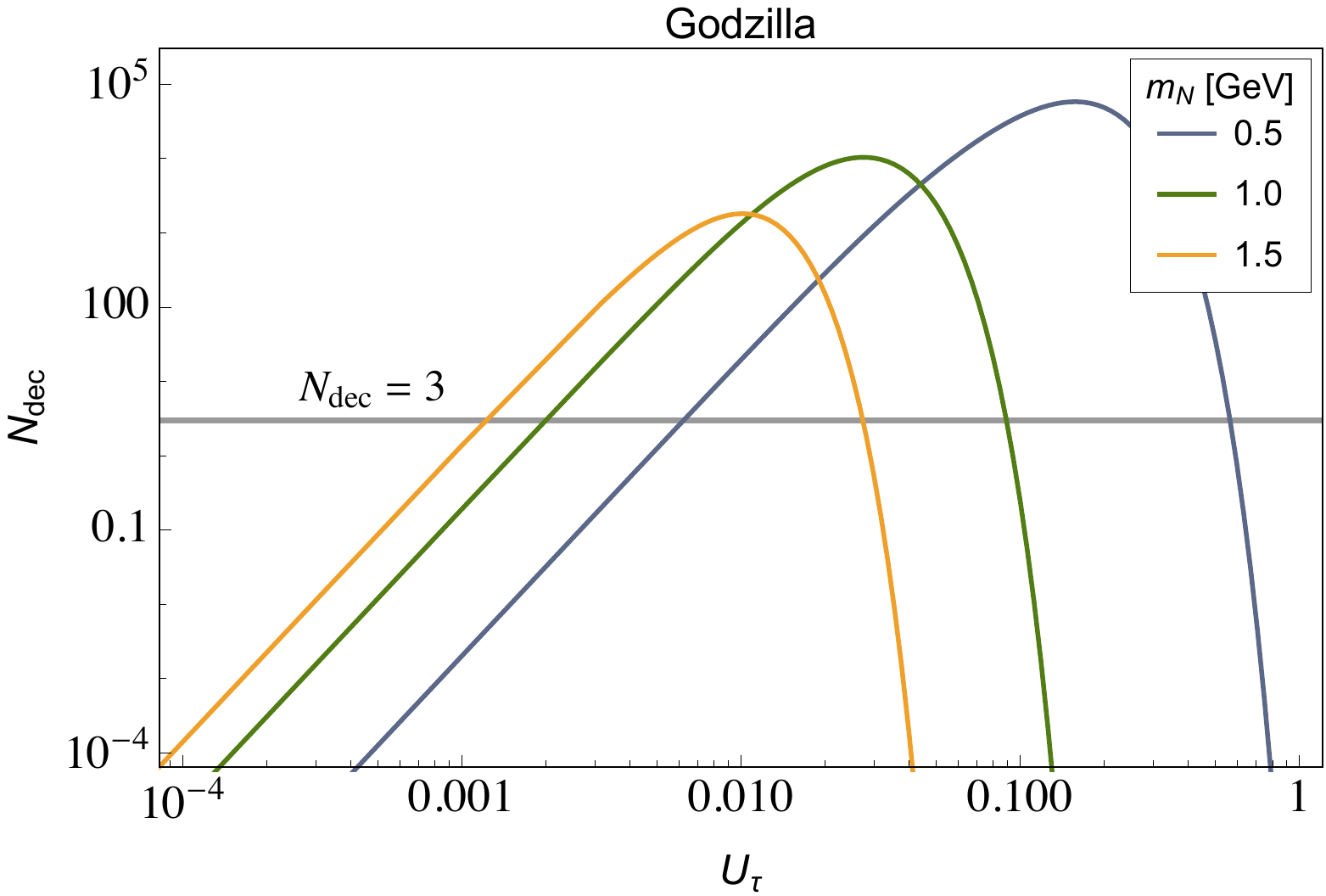}
\caption{Expected number of HNL events inside the GAZELLE detectors as a function of the mixing angle $U_\tau$, for three mass benchmarks.
}\label{fig:Nev:HNL} 
\end{center}
\end{figure}

In Fig.~\ref{fig:Nev:HNL}, we show the expected event yield at \gazelle, assuming 100\% efficiency and zero background in the detector.
%
In this best-case scenario, we can claim a signal with 95\% confidence over the null hypothesis if we observe at least $N_{\rm dec} = 3$ events. This value is represented by the horizontal line in Fig.~\ref{fig:Nev:HNL}. As an example, in this ideal case L-\gazelle would be able to probe mixing angles down to $U_\tau \sim 1.4 \times 10^{-3}$ for $m_N = 1.5$ GeV or $U_\tau \sim 7.1 \times 10^{-3}$ for $m_N = 0.5$ GeV.

These predictions need to be compared with the expected reach of \belletwo, assuming 100\% detector efficiency as for \gazelle. A previous study estimated that \belletwo could probe mixing angles down to $U_\tau \sim 1.6 \times 10^{-3}$ for $m_N = 1.5$ GeV and $U_\tau \sim 2 \times 10^{-3}$ for $m_N = 0.5$ GeV~\cite{Dib:2019tuj}.\footnote{The estimated detector efficiency in Ref.~\cite{Dib:2019tuj} is about $25 \%$, which we used to rescale the results to obtain the projections at $100 \%$ efficiency.} In Table~\ref{tab:HNLreach}, we show the comparison of the reach of L-\gazelle versus \belletwo for the three mass benchmarks. For high HNL masses, L-\gazelle is sensitive to smaller mixing angles $U_\tau$ than \belletwo, while the opposite is true for small masses. This strong mass dependence is due to the scaling of the HNL branching ratios. In particular, the three-body decay width of the HNL into charged states relevant for L-\gazelle scales as $m_N^5$, while \belletwo also probes other decay channels. This explains the larger gain of L-\gazelle over \belletwo for large HNL masses, despite a smaller average decay probability (Fig.~\ref{fig:prob:HNL}).

\begin{table}[t!]
\centering
 \begin{tabular}{|c c c c|} 
 \hline
 $m_N$ [GeV] & L-\gazelle & \belletwo 
 & LG/\belletwo \\ [0.5ex] 
 \hline\hline
 0.5 & $7.1 \times 10^{-3}$ & $2.0 \times 10^{-3}$ & 3.6 \\ 
 \hline
 1.0 & $2.2 \times 10^{-3}$ & $1.1 \times 10^{-3}$ & 2.0 \\
 \hline
 1.5 & $1.4 \times 10^{-3}$ & $1.6 \times 10^{-3}$ & 0.85 \\
 \hline
\end{tabular}
\caption{Projected reach of L-GAZELLE and \belletwo for the mixing angle $U_\tau$ with the three mass benchmarks considered. The last column shows the ratio of the reach at L-GAZELLE over \belletwo, assuming 100\% efficiency for both detectors. Ratios smaller than one indicate a better performance of L-GAZELLE.}
\label{tab:HNLreach}
\end{table}

The analysis performed here for $\tau$-flavored HNLs can be easily extended to the $e$- and $\mu$-flavored cases, $N_e$ and $N_\mu$ respectively. At \belletwo the main production mechanisms would be $\tau$, $D$ and $B$ decays. In the very light mass region $m_N \lesssim 100$ MeV, the HNL could be in principle produced with enough boost to largely escape the \belletwo detector. However, this region is already excluded by beam dump experiments, see {\it e.g.} Ref.~\cite{Beacham_2019}. For heavier HNLs, the gain of \gazelle over \belletwo would be dictated again by the respective fiducial coverage. 
\FloatBarrier

\subsection{Axion-like particles}\label{sec:ALPs}
\begin{figure}[t]
\begin{center}
\includegraphics[width=.5\linewidth]{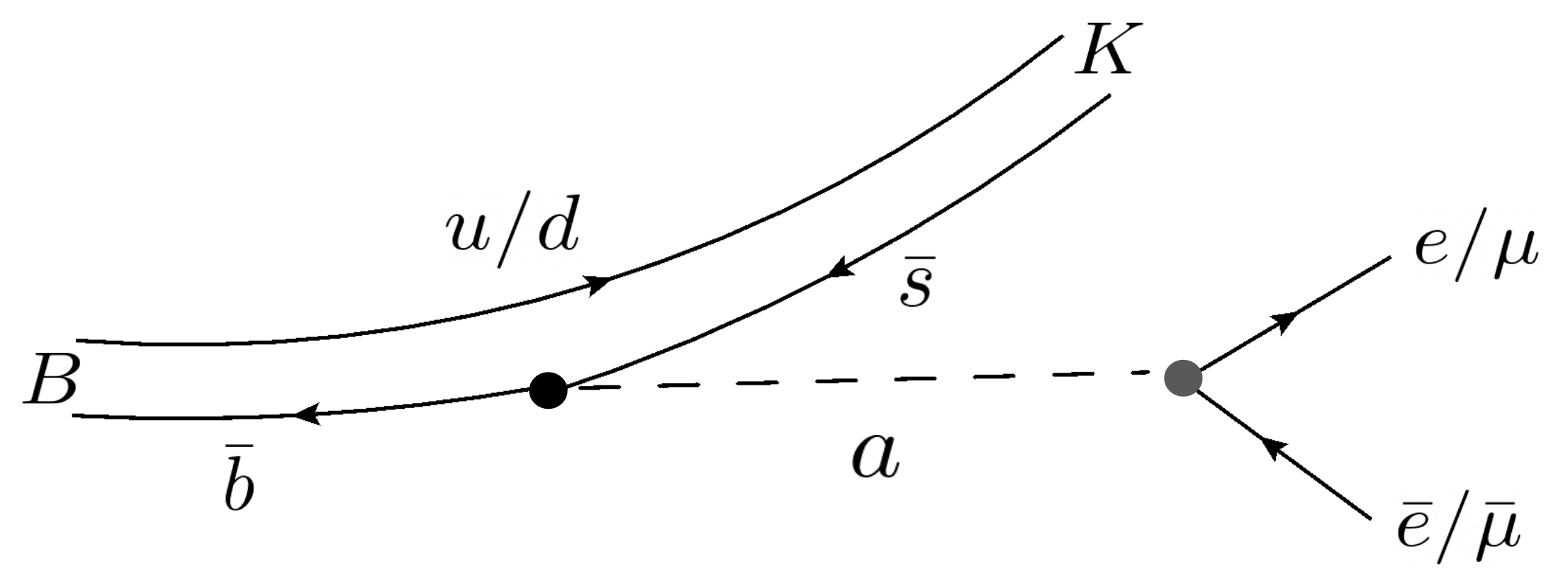}
\caption{Feynman diagram for ALP production via $B \to K a$ decays at~\belletwo.
}\label{fig:ALPs-feynman-diagram} 
\end{center}
\end{figure}

\noindent Axion-like particles (ALPs) arise in pseudoscalar extensions of the Standard Model. These hypothetical particles are pseudo Nambu-Goldstone bosons associated with a chiral symmetry, the Peccei-Quinn symmetry, that is broken at some high scale $\Lambda$~\cite{PhysRevLett.38.1440,PhysRevLett.40.223,PhysRevLett.40.279}. ALPs can couple to gauge bosons and fermions of the SM via dimension-5 operators (frequently referred to as the ``axion portal'' in the literature)~\cite{Jaeckel:2010ni,Ringwald:2014vqa,Marsh:2017hbv}. At \belletwo, ALPs with masses below the GeV scale can be abundantly produced as on-shell particles in $B$ meson decays. Subsequently, the ALPs travel through the detector and decay to leptonic or hadronic final states, if these are kinematically allowed. In our analysis, we focus on the final states with electron or muon pairs, $B \rightarrow K + a\, (a \rightarrow \ell^+ \ell^-)$ (Fig.~\ref{fig:ALPs-feynman-diagram}).

The relevant ALP interactions around the $B$ mass scale are described by an effective Lagrangian
\begin{equation}\label{eq:ALPs-lagrangian}
    \mathcal{L}_{\rm ALP} = -2 g_{i j} \frac{\partial_{\mu} a}{\Lambda}\,  \bar{d}_i \gamma^{\mu} P_L d_j + \frac{c_{\ell}}{2}\,\frac{\partial^{\mu} a}{\Lambda} \bar{\ell} \gamma_{\mu} \gamma_5 \ell,\qquad  i,j=1,2,3,\quad \ell = e,\mu,
\end{equation}
where $g_{ij}$ is the effective coupling of the ALP to down-quark FCNCs and $c_\ell$ is the coupling of ALPs to leptons.
The FCNC coupling, $g_{ij}$, can originate directly from a broken Peccei Quinn symmetry at a high scale $\Lambda$ or be generated from flavor-diagonal couplings in the electroweak theory at loop level~\cite{Bauer:2020jbp,Gavela:2019wzg}. For simplicity we assume that only $g_{23}=g_{sb}$ and $c_\ell=c_e=c_\mu$ are non-zero.\footnote{
 While the qualitative results of our study hold in general, the lifetime and leptonic branching ratios of the ALP can drastically change in the presence of additional couplings, such as couplings to photons, gluons, and light quarks.} The production rate of an on-shell ALP from $B$ decays is~\cite{Gavela:2019wzg}
    \begin{align}\label{eq:ALPs-prod-rate}
        \Gamma_{B\to K a} &= \frac{\left|g_{sb}\right|^2}{16\pi\Lambda^2} \left|f_0\left(m_a^2\right)\right|^2 \frac{\left(m_B^2-m_K^2\right)^2}{m_B}\left(1-\frac{\left(m_K+m_a\right)^2}{m_B^2}\right)^{\frac{1}{2}}\left(1-\frac{\left(m_K-m_a\right)^2}{m_B^2}\right)^{\frac{1}{2}},
    \end{align}
where $f_0\left(m_a^2\right)$ is the scalar hadronic form factor at
momentum transfer $q^2=m_a^2$, see {\it e.g.} Ref.~\cite{Gubernari:2018wyi}. In our numerical analysis, we consider only $B^+\to K^+ a$ decays. Owing to their nature as Nambu-Goldstone bosons, the decay rate of ALPs into leptons scales quadratically with the lepton mass~\cite{Bauer:2017ris,Bauer:2020jbp},
\begin{align}\label{eq:ALPs-dec-rate}
        \Gamma_{a\to\ell^-\ell^+} &= \frac{m_\ell^2\left|c_{\ell}\right|^2}{8\pi\Lambda^2} \sqrt{m_a^2-4m_\ell^2}.
\end{align}
For small ALP-lepton couplings $c_\ell$, the ALP decays at a distance from the production point. The signal resulting from $B^+$ decays consists of two charged lepton tracks that point to a displaced vertex and a charged kaon. The kaon momentum can be reconstructed from its decay products, allowing for an efficient background rejection.

\begin{figure}[h]
    \begin{center}
        \includegraphics[width=.575\linewidth]{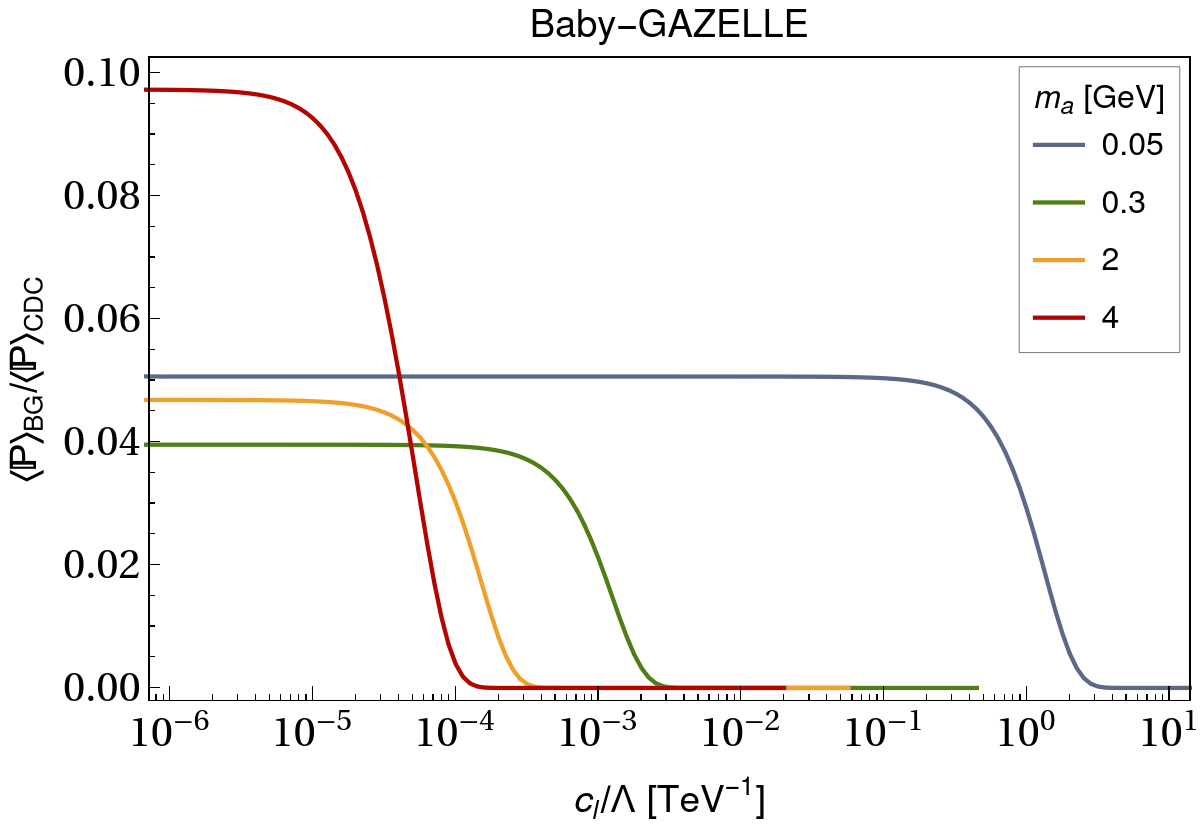}\\[0.95cm]
        \includegraphics[width=.575\linewidth]{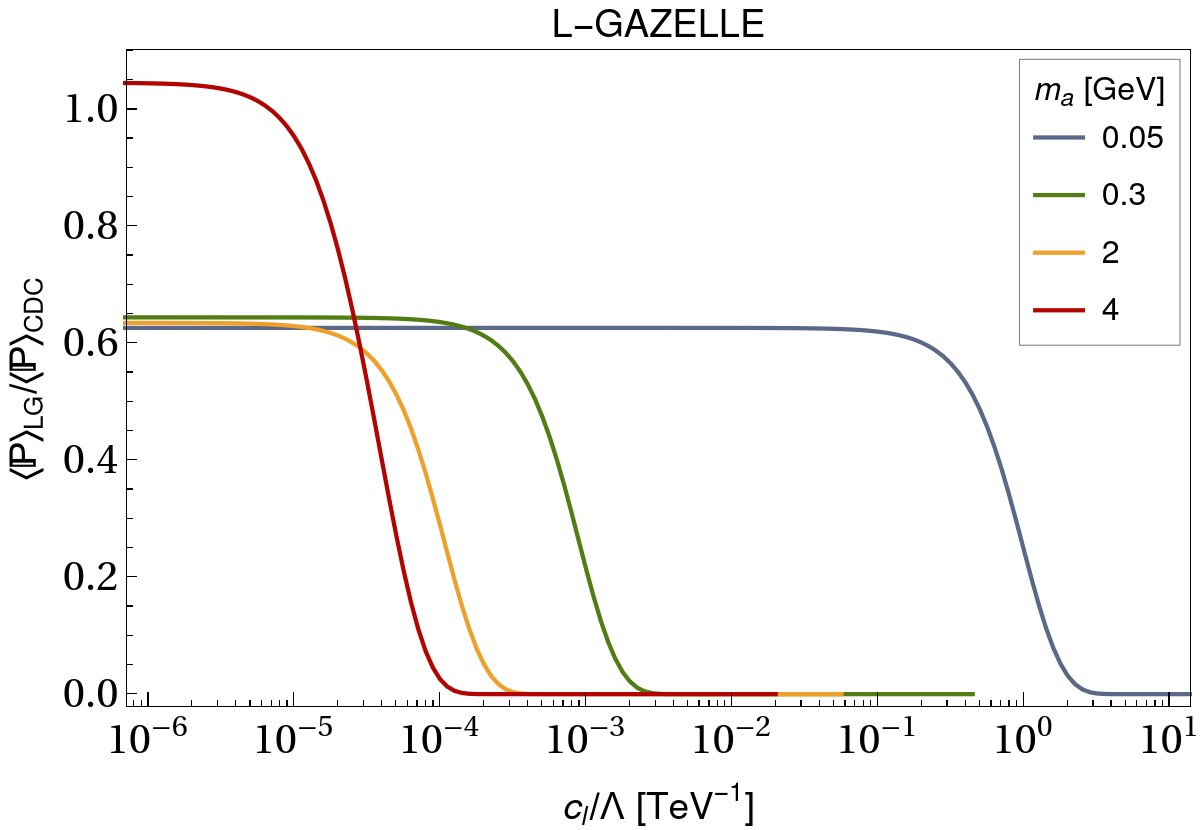}\\[0.95cm]
        \includegraphics[width=.575\linewidth]{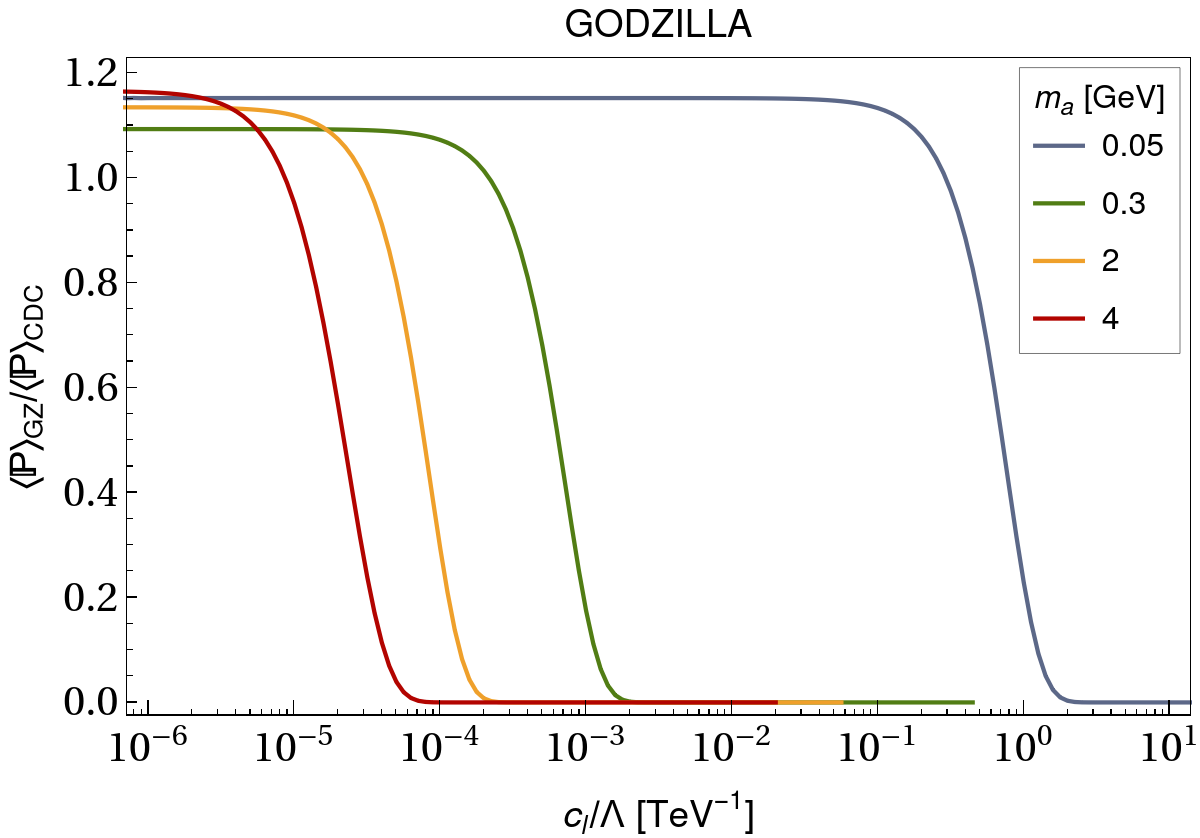}
        \caption{Ratio of average ALP decay probabilities within different configurations of \gazelle/\belletwo as a function of the effective ALP-lepton coupling $c_\ell/\Lambda$ and for various ALP masses $m_a$.}
        \label{fig:ALPs-prob-ratios} 
    \end{center}
\end{figure}

Following the procedure described in~Sec.~\ref{sec:models-intro} and using Eqs.~\eqref{eq:ALPs-prod-rate}~and~\eqref{eq:ALPs-dec-rate}, we calculate the probability of ALPs decaying inside the \gazelle configurations discussed in~Sec.~\ref{sec:detector} relative to \belletwo. The corresponding probability ratios for various ALP masses are shown in~Fig.~\ref{fig:ALPs-prob-ratios}.
 We can see that for small values of the coupling, which correspond to long lifetimes, \godzilla and (for some masses) L-\gazelle outperform \belletwo, while Baby-\gazelle has a much lower rate than \belletwo for all couplings.

At the value of $c_\ell/\Lambda$ for which the probability ratio rapidly drops in \belletwo's favor,  the ALP becomes short-lived on the scale of the \belletwo detector. We also notice that the sensitivity of Baby-GAZELLE and L-GAZELLE is significantly better for heavy ALPs with masses near the kinematic endpoint in $B\to K$ decays. The reason is that heavy ALPs inherit
the forward boost of the $B$ mesons, so that a far-distance detector positioned in the forward region captures more of them.

\begin{figure}[t]
\begin{center}
\includegraphics[width=.575\linewidth]{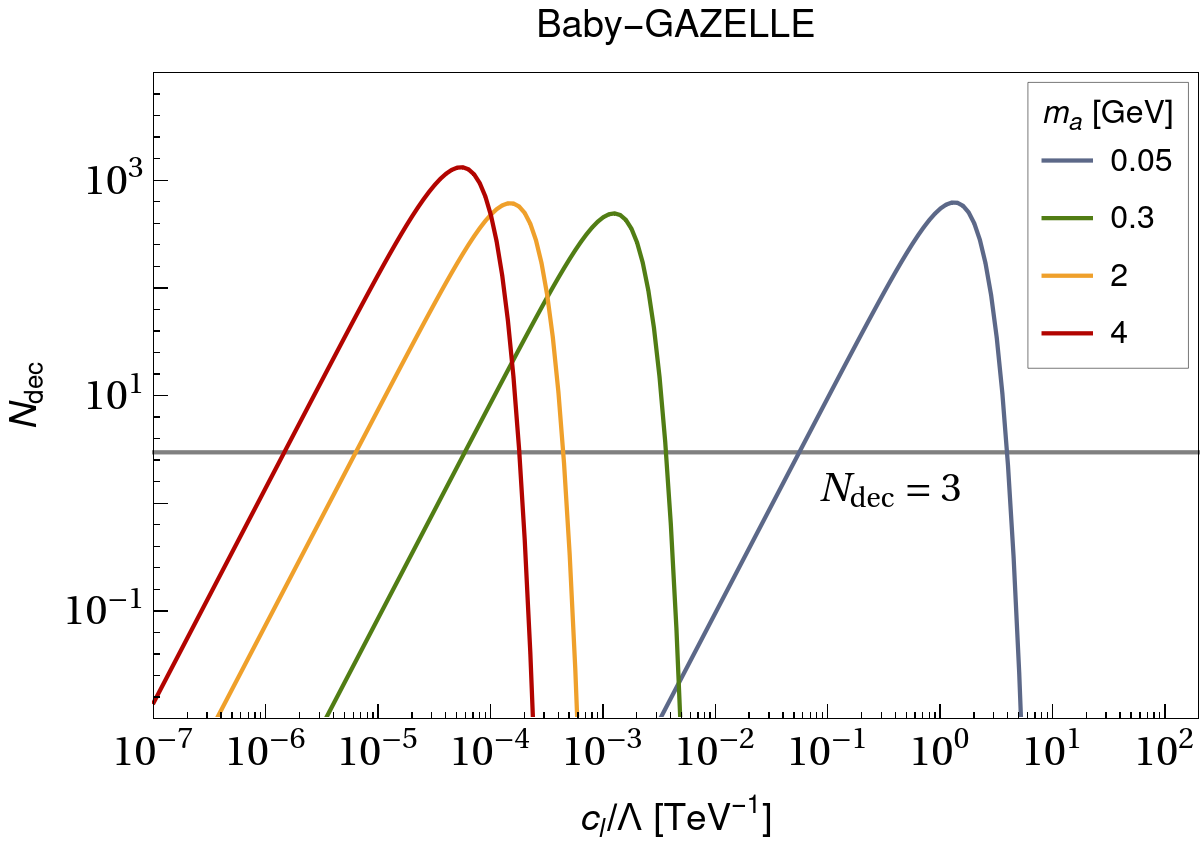}\\[1cm]
\includegraphics[width=.575\linewidth]{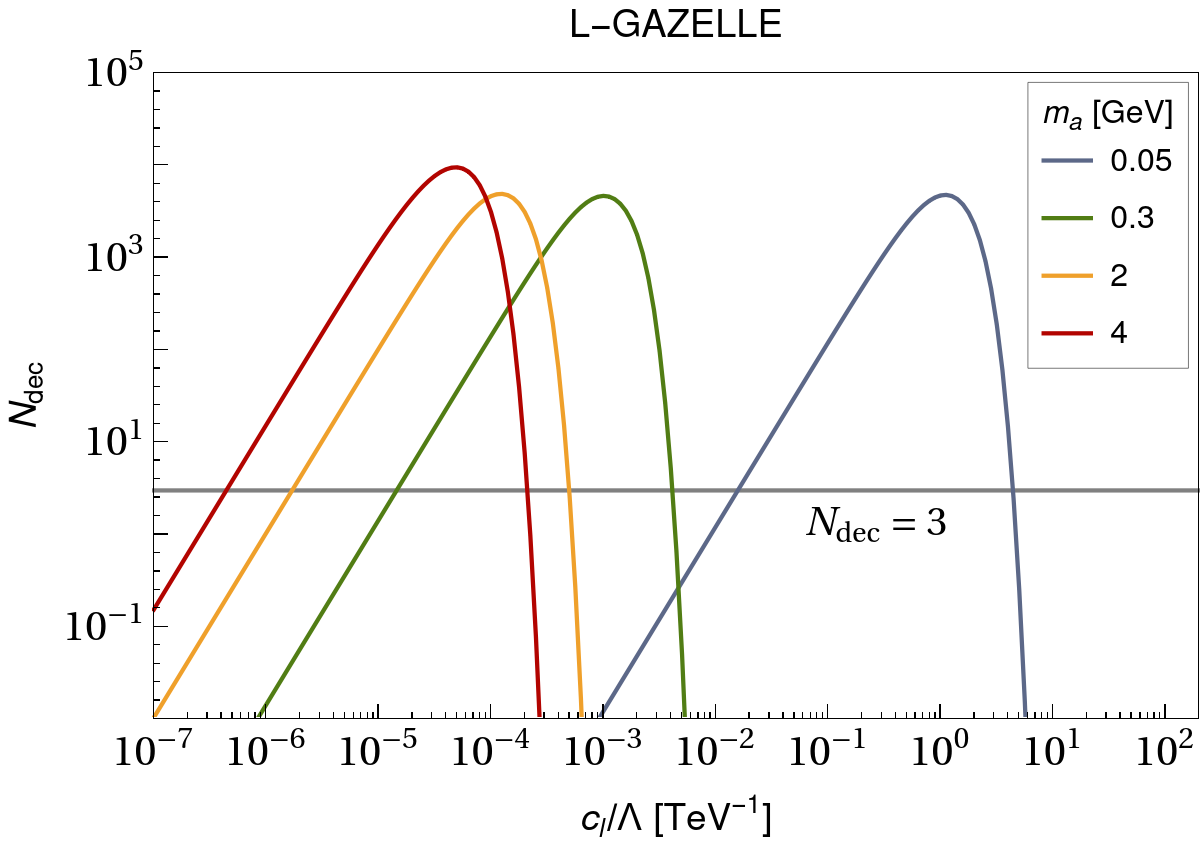}\\[1cm]
\includegraphics[width=.575\linewidth]{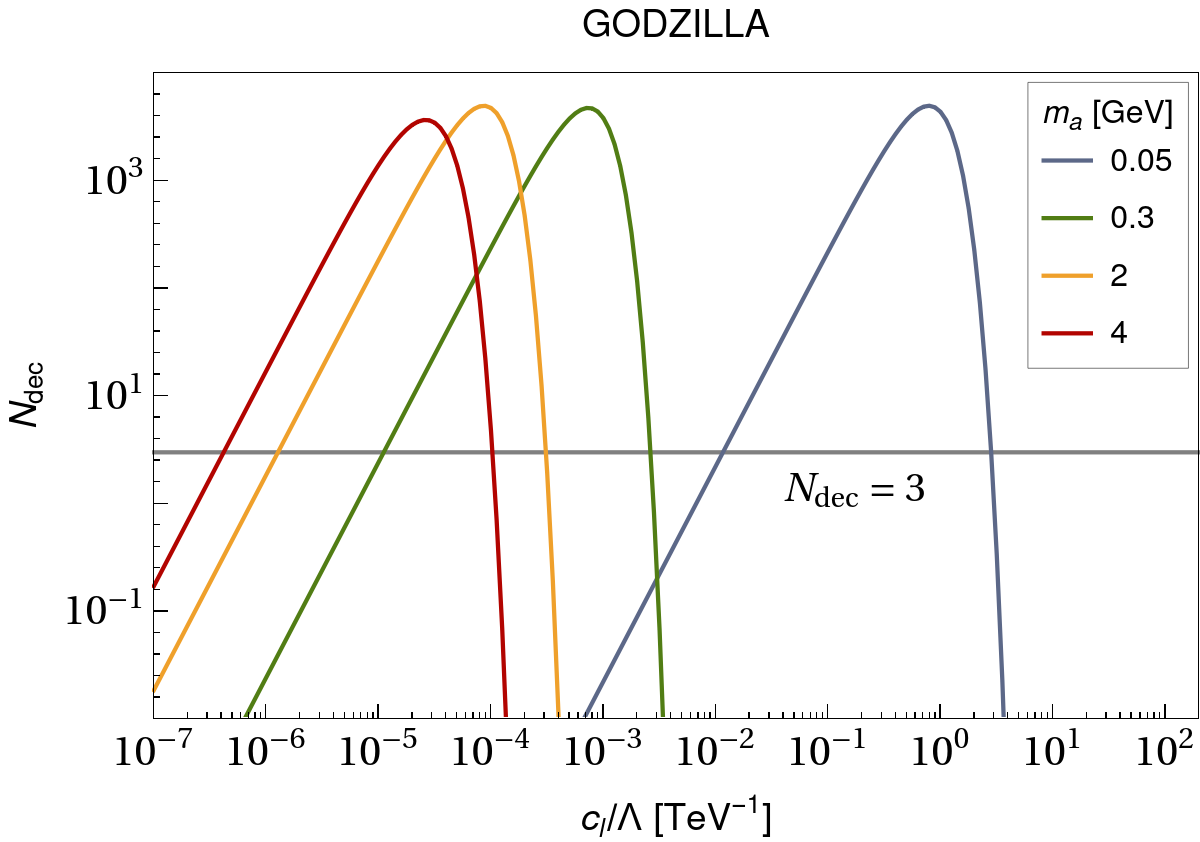}
\caption{Maximum expected number of events in the \gazelle detectors as a function of the ALP-lepton coupling $c_\ell/\Lambda$ for various ALP masses.}
\label{fig:ALPs-events} 
\end{center}
\end{figure}

We also estimate the maximum number of events expected at~\gazelle,
\begin{align}
    N_{\rm dec} = N_{B\bar{B}} \times \mathrm{Br}_{B^\pm\to K^\pm a}(m_a,g_{sb}) \times \mathrm{Br}_{a\to \ell^+\ell^-}(m_a,c_\ell) \times \langle\mathds{P}(m_a,c_\ell,g_{sb})\rangle,
\end{align}
where $N_{B\bar{B}} = 5 \times 10^{10}$ corresponds to the number of $B\bar{B}$ pairs produced at Belle II with $50$ ab$^{-1}$ luminosity. For the ALP production rate, we determine the largest coupling $g_{sb}$ that is allowed by previous searches at flavor experiments for $B\to K$ 
 decays with missing energy in the final state, assuming that the majority of LLPs escape the detector. To date, the strongest upper bounds on 
$g_{sb}$ come from searches for
 $B \rightarrow K + {\rm invisible}$ at BaBar~\cite{Lees:2013kla}.
 In our numerical analysis we take these bounds into account and fix $g_{sb}$ to the values given in Table~\ref{tab:ALPsreach}.
\begin{table}
\centering
 \begin{tabular}{|c c c c c|} 
 \hline
 $m_a$ [GeV] & $g_{sb}$ & L-\gazelle & \belletwo & LG/\belletwo \\ [0.5ex] 
 \hline\hline
 0.3 & $3.9\times 10^{-6}$ & $1.4\times10^{-5}$ & $1.2\times10^{-5}$ & 1.2 \\ 
 \hline
 2.0 & $3.8\times 10^{-6}$ & $1.7\times10^{-6}$ & $1.4\times10^{-6}$ & 1.3 \\
 \hline
 4.0 & $3.5\times 10^{-6}$ & $4.4\times10^{-7}$ & $4.5\times10^{-7}$ & 1.0 \\
 \hline
\end{tabular}
\caption{Projected reach of L-\gazelle and \belletwo for the ALP coupling $c_\ell/\Lambda\,[\text{TeV}^{-1}]$ for three mass benchmarks. The second line shows the maximum coupling $g_{sb}$ allowed by current bounds from flavor experiments (see text). The last column shows the ratio of the reach at L-\gazelle over \belletwo, assuming 100\% detection efficiency. Ratios smaller than one indicate a better performance of L-GAZELLE.}
\label{tab:ALPsreach}
\end{table}
 The resulting number of $\ell^+ \ell^-$ pairs produced in the \gazelle detector is shown in Fig.~\ref{fig:ALPs-events}.
 As in the previous section, we mark the line of $N_{\rm dec}=3$ events above which we can claim a signal with 95\% CL given zero background.

In Table~\ref{tab:ALPsreach}, we show the smallest ALP couplings $c_\ell/\Lambda$ to which L-\gazelle and \belletwo are sensitive if a 100\% detection efficiency is assumed. 
While the reach is comparable to \belletwo's, it is a bit smaller, which we discuss more generally in Sec.~\ref{sec:generalaspects}.
The sensitivity gain of L-\gazelle increases with the ALP mass, because heavy ALPs are more abundantly produced in the forward region and missed by \belletwo.

A detailed study of an inclusive search for $B \rightarrow K + {\rm invisible}$ at \belletwo can be found in \cite{Ferber:2022rsf}.
\FloatBarrier

\subsection{Inelastic dark matter}\label{sec:IDM}
\noindent Thermal dark matter of a few GeV is severely constrained by CMB observations and direct detection searches~\cite{TuckerSmith:2001hy}. These constraints can be avoided if dark matter couples inelastically to the SM fields.  As discussed in, {\it e.g.}, Refs.~\cite{Izaguirre:2015zva,Izaguirre:2017bqb,Berlin:2018jbm,Duerr:2019dmv}, the most straightforward realisation of this idea, dubbed inelastic dark matter (iDM), is a pair of Majorana fermions, $\chi_1$ and $\chi_2$, coupled to a massive dark gauge boson, $A'$, which kinetically mixes with the Standard Model. In this scenario, $\chi_1$ is the dark matter candidate, $m_{\chi_2}> m_{\chi_1}$ to avoid the stringent direct detection limits resulting from $A'$ exchange with nuclear targets, and $m_{A'}> m_{\chi_1}$ to avoid CMB limits associated with the annihilation $\chi_1\chi_1 \to A'A'$ in the early universe. A natural setup to explain the $A'$ mass, as well as the mass splitting between $\chi_1$ and $\chi_2$, is to introduce a Higgs mechanism in the dark sector~\cite{Duerr:2020muu}. 
We will adopt this iDM setup and study the production and decay of the associated dark scalar $h'$.
The dark sector has seven free parameters: the masses associated with $A'$, $h'$, $\chi_1$ and $\chi_2$, the angle $\theta$ characterizing the mixing with the SM scalar, the kinetic mixing $\epsilon$, and the dark fine structure constant $\alpha_D$. We refer the reader to Ref.~\cite{Duerr:2020muu} for details.

\begin{figure*}[h!]
\begin{center}
\includegraphics[width=0.6\textwidth]{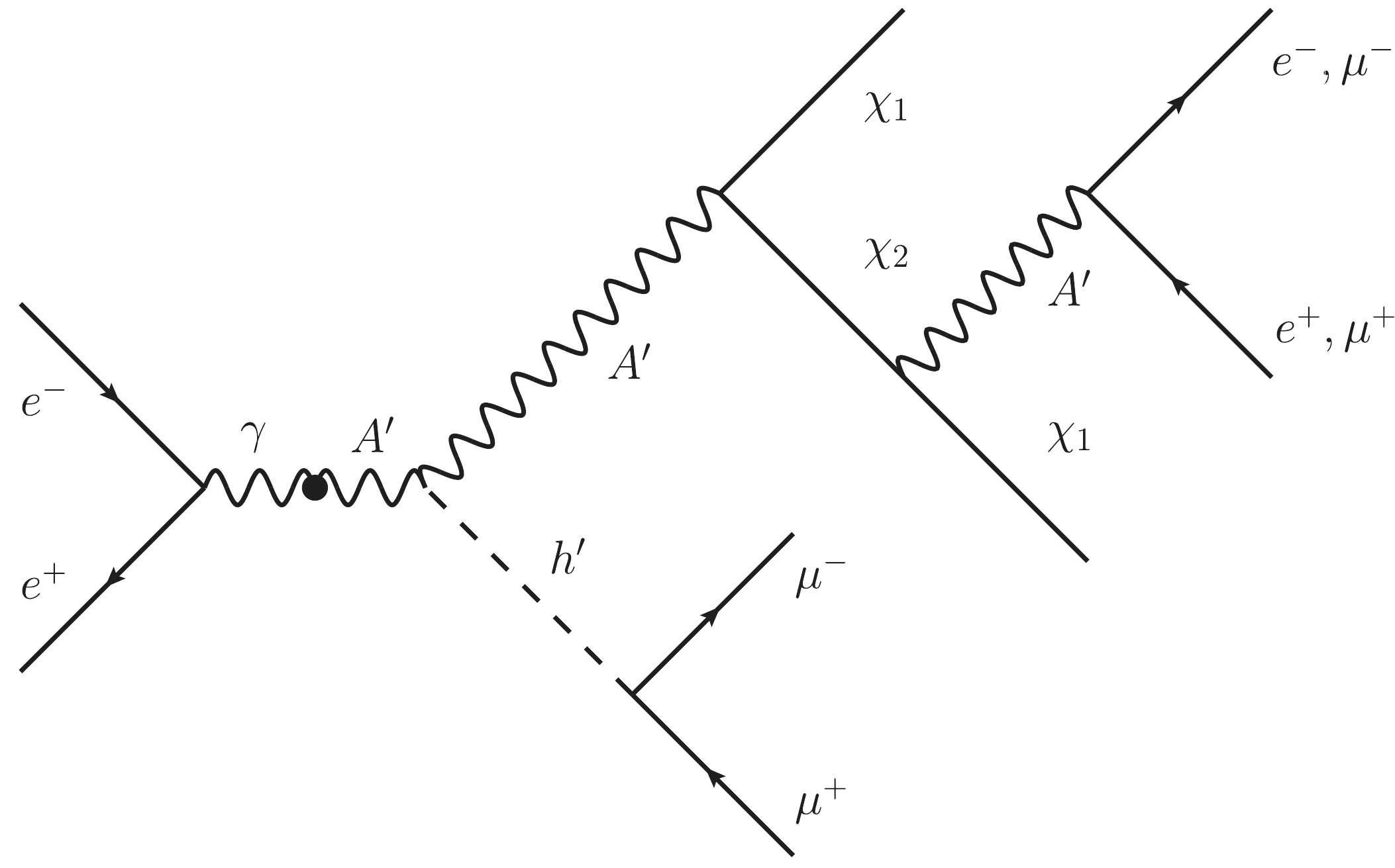}
\caption{Feynman diagram for dark Higgs production and decay in the iDM model.}
\label{fig:figidm}
\end{center}
\end{figure*}

At \belletwo the dark Higgs can be copiously produced if it is radiated from a dark photon, as indicated in Fig.~\ref{fig:figidm}. To assess the prospects of detecting such dark scalars in the different configurations of \gazelle, we will assume that they decay into a pair of muons.  

 Accordingly, we simulate the process $e^+ e^- \to \chi_1 \chi_1 \ell^+ \ell^- h'$ in \texttt{MadGraph}\,\cite{Alwall:2014hca}, where $\ell=e$ or $\mu$. The corresponding  cross section does not depend on the mixing angle and can be large if the dark photon produced in association with $h'$ is on-shell. In this case the cross section can be computed from 
\begin{equation}
\frac{d\sigma(e^+e^-\to A' h')}{d\cos \vartheta}=
\frac{8 \pi  \alpha  \alpha_D \epsilon ^2 }{\sqrt{s} \left(s-m_{A'}^2\right)^2}\, k\left(2 m_{A'}^2+k^2 \sin^2\vartheta \right),
\end{equation}
where $\sqrt{s}$ is the center-of-mass energy at \belletwo,
\begin{align}
    k= \frac{\sqrt{s}}{2}\left(1-\frac{(m_{A'}-m_{h'})^2}{s}\right)^{\frac{1}{2}}\left(1-\frac{(m_{A'}+m_{h'})^2}{s}\right)^{\frac{1}{2}}
\end{align}
is the three-momentum of the dark Higgs in the center-of-mass frame, and $\vartheta$ is the corresponding scattering angle.  
Motivated by current experimental limits~\cite{Duerr:2020muu}, we fix the following parameters
\begin{align}
\epsilon=10^{-3}\,,&&
\alpha_D=0.1\,,&&
m_{A'}=8~\text{GeV}\,,&&
m_{\chi_1}= 2~\text{GeV}\,,&&
m_{\chi_2}= 4~\text{GeV}.
\label{eq:idmparset}
\end{align}
We will consider three benchmarks for the dark Higgs mass, $m_{h'}= 0.4,\,1, \mathrm{and\,} 3$~GeV. The expected number of events from the decay of the dark Higgs in the different \gazelle configurations is evaluated through 
\begin{align}
    N_{\rm dec} = N_{\chi_1 \chi_1 \ell^+ \ell^- h'} \times \mathrm{Br}_{h'\to \mu^+\mu^-}\left(m_{h'}\right) \times \langle\mathds{P}(m_{h'},\theta)\rangle\,.
\end{align}
The number of iDM events $N_{\chi_1 \chi_1 \ell^+ \ell^- h'}$ is calculated from the cross section obtained by \texttt{MadGraph} assuming 50 ab$^{-1}$ of data. The branching ratio for $h'\to \mu^+\mu^-$ is evaluated using
\begin{align}\label{eq:mudecay}
\hspace{-0.3cm} \mathrm{Br}_{h'\to \mu^+\mu^-}\left(m_{h'}\right)  = \frac{\Gamma_{h'\to\mu^+\mu^-}}{\Gamma_\text{tot}},&& \text{with}&&\hspace{-0.1cm} \Gamma_{h'\to\mu^+\mu^-}=\frac{\sin^2\theta\,G_F\,m_{h'}\, m_\mu^2}{4\sqrt{2}\,\pi}\,\left(1-\frac{4m_\mu^2}{m_{h'}^2}\right)^{\frac{3}{2}}\hspace{-0.1cm}.
\end{align}
In order to account for the hadronic contributions to the total width of the dark scalar, $\Gamma_\text{tot}$, we employ the results from Ref.~\cite{Winkler:2018qyg}. 
Finally, the decay probability $\langle\mathds{P}(m_{h'},\theta)\rangle$ is computed following the method described around Eqs.~\eqref{eq:LLP:probability} and \eqref{eq:averageprob}. We note that in the mass range of interest the dark Higgs width $\Gamma_\text{tot}$ is proportional to $\sin^2\theta$. Therefore the decay probability depends on the mixing angle through the lifetime of the dark Higgs, $\tau_{h'} = 1/\Gamma_{\rm tot}$, while the branching ratio $\mathrm{Br}_{h'\to \mu^+\mu^-}$ is independent of $\theta$ and only depends on $m_{h'}$.

\begin{figure*}[t]
\begin{center}
\includegraphics[width=.575\linewidth]{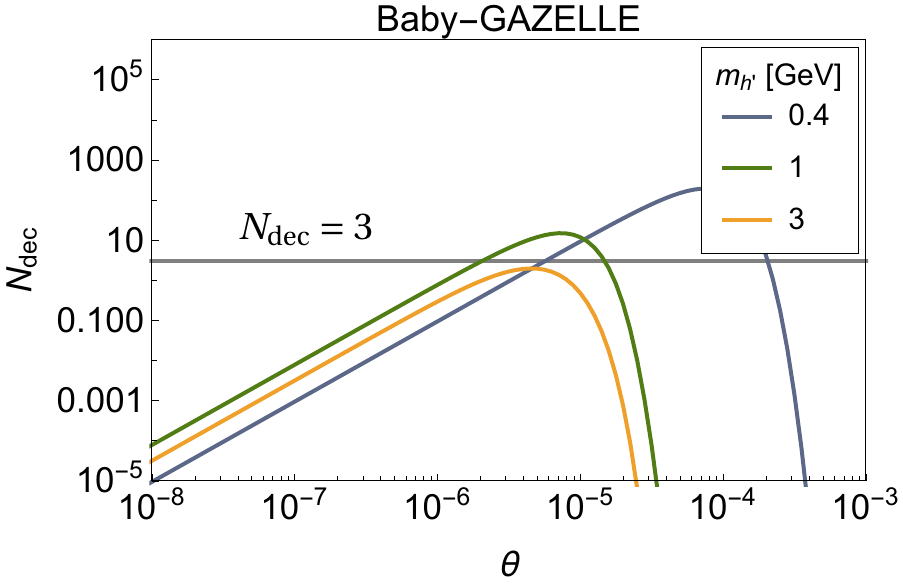}\\[1cm]
\includegraphics[width=.575\linewidth]{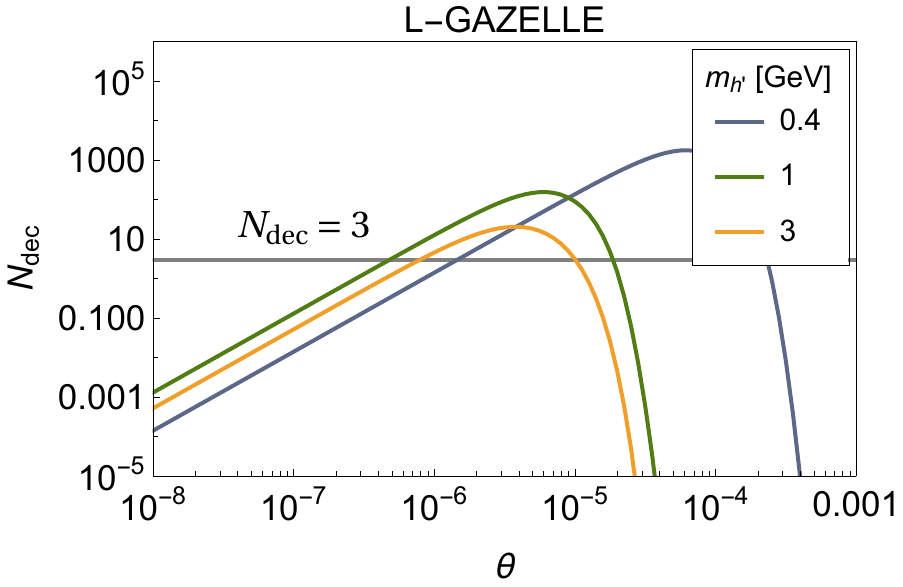}\\[1cm]
\includegraphics[width=.575\linewidth]{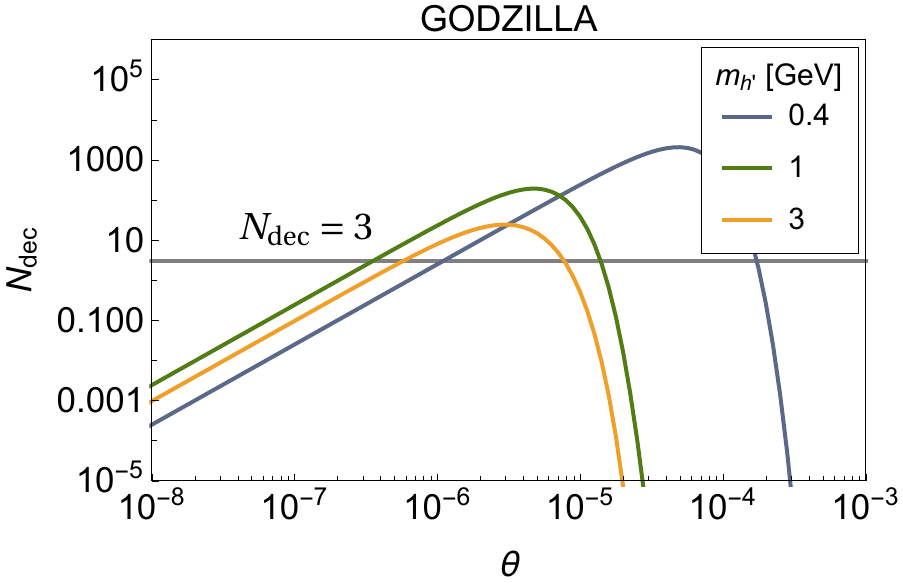}
\caption{Expected number of dark Higgs events from iDM as a function of the scalar mixing angle, for the parameter set
 in Eq.~\eqref{eq:idmparset} and the indicated dark Higgs masses.}
\label{fig:events_idm}
\end{center}
\end{figure*}

\begin{figure*}[t]
\begin{center}
\includegraphics[width=.575\linewidth]{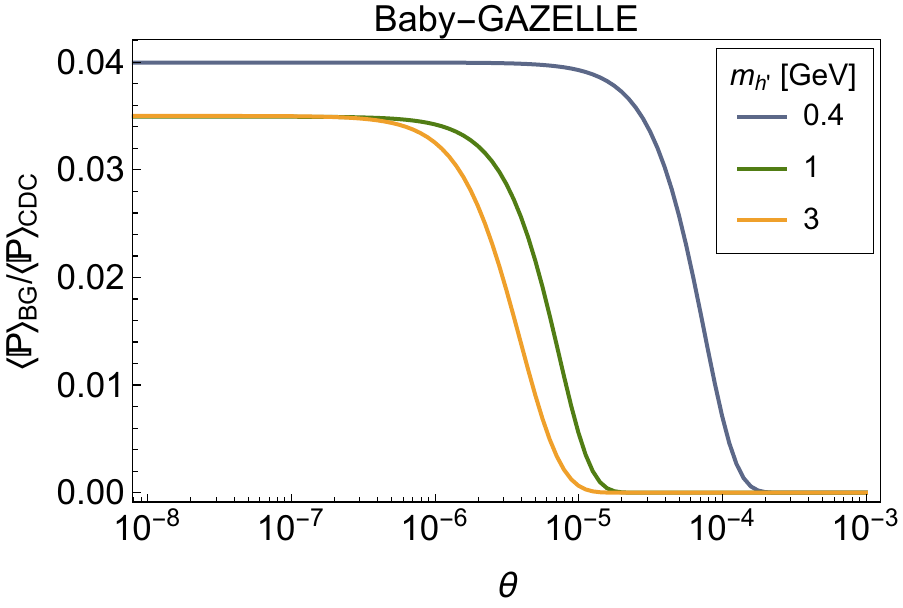}\\[1cm]
\includegraphics[width=.575\linewidth]{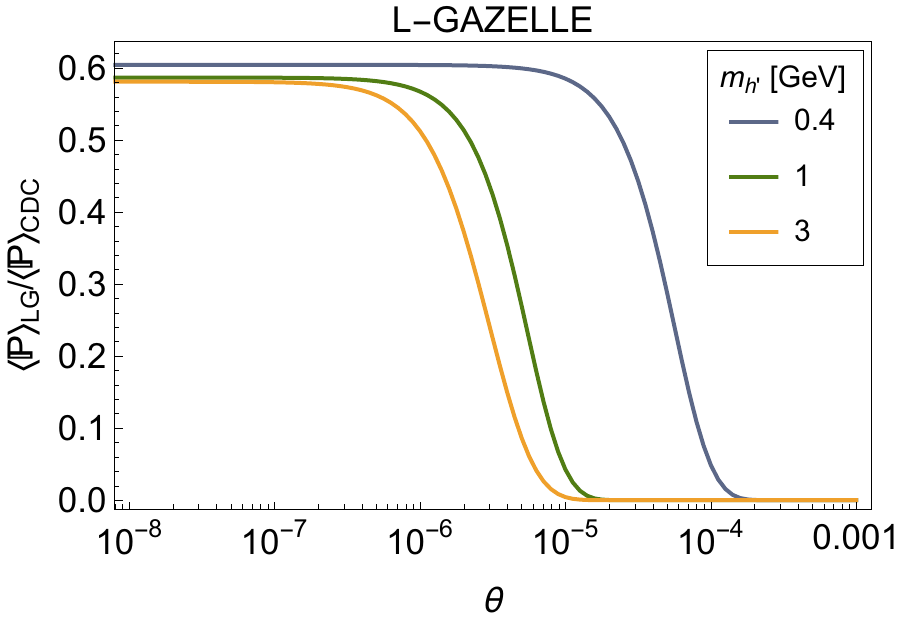}\\[1cm]
\includegraphics[width=.575\linewidth]{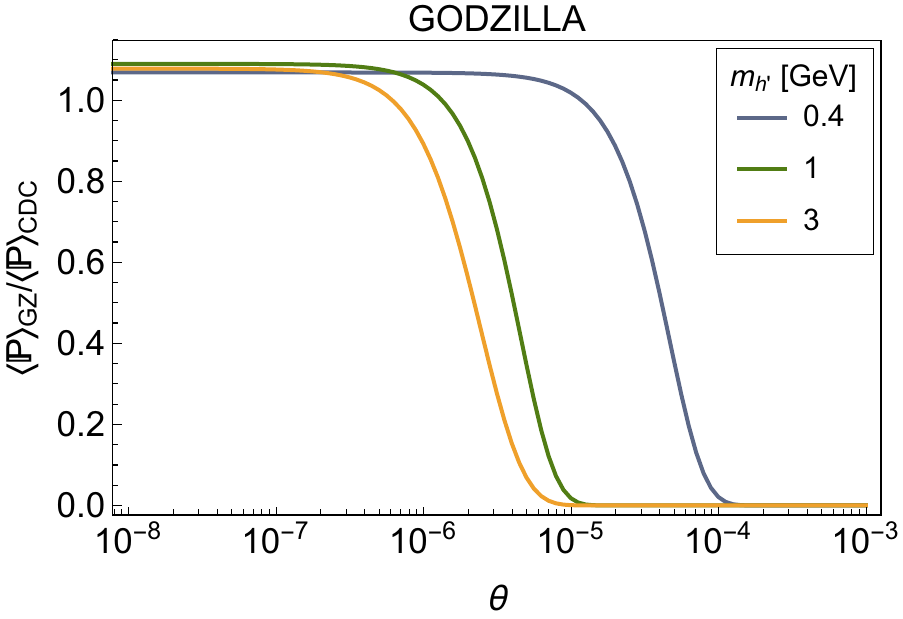}
\caption{Ratio of average decay probabilities for a dark Higgs in the iDM model within different configurations of \gazelle/\belletwo, as a function of the scalar mixing angle $\theta$, for the parameter set in Eq.~\eqref{eq:idmparset} and the indicated dark Higgs masses.}
\label{fig:probabilities_idm}
\end{center}
\end{figure*}

In Fig.~\ref{fig:events_idm}, we show the expected number of decay events as a function of the scalar mixing angle $\theta$ for the three scalar masses mentioned above.
Similarly, in Fig.~\ref{fig:probabilities_idm} we compare the probabilities of detecting a dark Higgs decay in the different configurations of \gazelle against those for \belletwo. 
 For $\theta \gtrsim 10^{-4}$, \belletwo can detect many more events, whereas for smaller angles the detection probabilities in the \gazelle detectors are similar to those at \belletwo, especially for GODZILLA and L-\gazelle. In Table~\ref{tab:idm_reach}, we show the smallest $\theta$ to which L-\gazelle and \belletwo are sensitive if a 100\% detection efficiency is assumed. For all benchmarks, their ratio is only marginally bigger than 1.
%
    
\begin{table}[t!]
\centering
\begin{tabular}{|c c c c|}
 \hline
 $m_{h'}$ [GeV] & L-\gazelle & \belletwo
 & LG/\belletwo \\ [0.5ex]
 \hline\hline
 0.4 & $1.46 \times 10^{-6}$& $1.14 \times 10^{-6}$  & 1.29 \\
 \hline
 1.0 & $4.82 \times 10^{-7}$& $3.7 \times 10^{-7}$ & 1.31 \\
 \hline
 3.0 & $7.93 \times 10^{-7}$ & $5.8 \times 10^{-7}$ & 1.37  \\
 \hline
\end{tabular}
\caption{Projected reach of L-GAZELLE and \belletwo for the mixing angle $\theta$ with the three mass benchmarks considered. The last column shows the ratio of the reach at L-GAZELLE over \belletwo, assuming 100\% efficiency for both detectors. Ratios smaller than one indicate a better performance of L-GAZELLE.}
\label{tab:idm_reach}
\end{table}

\FloatBarrier


\section{Comparing the LLP reach at \gazelle and \belletwo}\label{sec:generalaspects}
\noindent 
Despite the different LLP production and decay modes, the overall sensitivity of GAZELLE does not significantly exceed the one of \belletwo in most of the benchmarks we considered. The high sensitivity of \belletwo and the comparably modest additional reach of \gazelle is mostly due to the geometry and the positioning of the detectors.

The sensitivity to LLPs with large decay lengths, $d\equiv\gamma\beta c\tau$, is mostly determined by the solid angle $\Omega$ (as viewed from the collision point) times the radial thickness $D$ of the detector, see also Sec.~\ref{sec:intro}. This follows from rewriting 
 the radial decay probability, Eq.~\eqref{eq:LLP:probability}, as
\begin{align}
    \mathds{P}(\ell^{in}) = \exp\left(-\frac{\ell^{in}}{d}\right) - \exp\left(-\frac{\ell^{in} + D}{d}\right),
\end{align}
and expanding the exponential for $d \gg \ell^{in},D$,  in which case the average decay probability is approximated by
\begin{align}\label{eq:P:parametric}
     \langle\mathds{P}\rangle \approx \Omega \times \frac{D}{d}\,.
\end{align}
Using the geometry parameters from Sec.~\ref{sec:detector}, we can roughly estimate the fiducial acceptance $\Omega \times D$ for \belletwo and the three \gazelle detectors as
\beq
\label{eq:acceptance}
\begin{matrix*}[l]
\text{Belle II}:\quad & 11.5\,\mathrm{sr} \times 0.6\,\mathrm{m} \commentout{11.5 \text{ sr } \times\ 0.6\text{ m }} & = \ 6.9\,\mathrm{sr}\,\mathrm{m},\\
\text{Baby-\gazelle}:\quad & 0.12\,\mathrm{sr}\times2.5\,\mathrm{m}\commentout{0.05 \text{ sr } \times\ 4\text{ m }} & =  \ 0.3\,\mathrm{sr}\,\mathrm{m}\commentout{0.2\text{ sr m}},\\
\text{L-\gazelle}:\quad & 0.34\,\mathrm{sr} \times 5.3\,\mathrm{m} + 0.76\,\mathrm{sr} \times 4.1\,\mathrm{m} \commentout{0.34 \text{ sr } \times\ 6\text{ m } +\  0.32 \text{ sr } \times\ 3\text{ m }} & =  \ 4.9\,\mathrm{sr}\,\mathrm{m}\commentout{3\text{ sr m}},\\
\text{GODZILLA}:\quad & 0.76\,\mathrm{sr} \times 4.1\,\mathrm{m} \commentout{0.17 \text{ sr } \times\ 20\text{ m }} & =  \ 8.9\,\mathrm{sr}\,\mathrm{m}\commentout{3.4 \text{ sr m}}.
\end{matrix*}
\eeq
For Baby-\gazelle, the fiducial acceptance is too small to compete with \belletwo. For L-\gazelle and GODZILLA, the fiducial acceptance is comparable to that of \belletwo, resulting in a similar sensitivity to LLPs in the limit of long decay lengths. This explains the typical decay ratios of $\langle \mathds{P}\rangle_{\text{LG/GZ}}/\langle \mathds{P}\rangle_{\text{CDC}} \approx 1$ for the three models in Figures~\ref{fig:prob:HNL}, \ref{fig:ALPs-prob-ratios} and \ref{fig:probabilities_idm} at small couplings. In all three models the reach at long lifetimes is determined by the detector geometry, rather than the exponential decay probability, which dominates for smaller decay lengths. Deviations from the simple estimates in Eq.~\eqref{eq:acceptance} are due to the kinematic distribution of the LLPs, as well as the limited validity of approximation \ref{eq:P:parametric}.

At future $e^+ e^-$ colliders, geometric limitations can be overcome by optimizing the fiducial acceptance. For instance, this idea has been put forward in the HADES proposal~\cite{Chrzaszcz:2020emg}, a $4\pi$ detector positioned at a distance around the collision point of the planned future electron-positron collider FCC-ee.

Finally, we comment on the complementarity of displaced vertex and missing energy searches. Due to its high angular coverage, \belletwo can efficiently detect LLPs with very long lifetimes as missing energy. For example, in the ALP model from Sec.~\ref{sec:ALPs} the projected rates for ALP decays in \gazelle are limited by existing searches for $B\to K + \rm{invisible}$ at BaBar\,\cite{Lees:2013kla}. Searches with displaced vertices or missing energy searches are complementary, as they cover all possible decay channels of neutral LLPs.
 In general, the sensitivity to signatures with missing energy depends on the detector efficiency to reconstruct the remainder of the event and on the backgrounds that can mimic the signal.
 Compared with the LHC experiments, \belletwo has the advantage of lower backgrounds, and compared to LHCb also the advantage of a nearly full angular coverage. On the other hand, event rates at \belletwo are typically lower than at the LHC.
 The reach obtained by either displaced vertices or missing energy searches, at the LHC or at \belletwo, depends on the details of the new physics models, and dedicated comparisons are warranted on a case by case basis. 

\section{Forward physics and confining dark sectors}\label{sec:exotica}
The three benchmark models discussed in Section~\ref{sec:models-intro} are representative of a broad range of LLPs that would typically arise in weakly coupled hidden sectors. 
 In this section we discuss several new physics scenarios that go beyond our previous discussion. In Section~\ref{sec:forward}, we explore the gain of \gazelle to detect light LLPs that are produced in the forward direction, and thus tend to escape the \belletwo detector acceptance. In Sections~\ref{sec:emjets}, \ref{sec:quirks}, and \ref{sec:soft:bombs} we then discuss three possible signatures of hidden sectors that contain a confining dark strong force. The confinement may lead to novel phenomena such as emerging jets, quirks, and soft bombs that could be seen at a far detector such as \gazelle.

\subsection{Forward LLP production}\label{sec:forward}
Belle\,II's lack of angular coverage in the forward region reduces the sensitivity to LLPs produced along the beam axis. If LLPs are predominantly produced in the forward direction, \gazelle would detect the LLP decay products with an improved efficiency compared to \belletwo. 

 An example of this effects is the ALP benchmark with mass $m_a = 4\,$GeV, close to the kinematic threshold in $B \to K a$ decays, see Section~\ref{sec:ALPs}. The forward boost of the $B$ meson at \belletwo is inherited by the ALP, which is then also emitted in the forward direction and as such tends to escape the Belle II detector. A far-distance detector positioned in the forward region would have an enhanced probability to detect such heavy LLPs.

Another example of a forward process is the production of a dark photon $A'$ via $e^+e^-\to \gamma A'$. This process features a collinear enhancement, even if the dark photon is moderately heavy, see, \emph{e.g.},  Eq.~(7) in Ref.~\cite{An:2015pva}, as well as the discussion in Ref.~\cite{Chen:2020bok}. The logarithmic divergence in the forward direction is cut off by the mass of the incoming electrons \cite{Fayet:2007ua}. 
Based on a rough estimate, we find three times as many dark photons produced in the forward direction compared to the event rates within the polar angle acceptance of \belletwo in the center-of-mass frame, $0.22 \lesssim \vartheta^{*} \lesssim 2.42$. This may lead to some sensitivity improvement at a far detector in certain models. In the minimal scenario where $A'$ only couples to electrons, this is not the case since $A'$ would be too short-lived to reach the far detector before decaying\,\cite{Ferber:2022ewf}. However, in the non-minimal scenarios from Section~\ref{sec:emjets}, where the $A'$ decays to dark sector states, which then decay back to the SM particles far from the interaction point, a forward detector could have an improved discovery reach over \belletwo.

We believe that similar arguments will apply to other NP setups such as a light axial vector, scalar or pseudo-scalar state that couples to electrons.
A simple immediate example for a pseudo-scalar is an ALP with couplings to electrons,\footnote{Setting aside the  model building challenge posed by the fact that, if this is a true PNGB, the interactions to electrons are suppressed by the electron mass and thus the phenomenologically relevant scale $f$ would be very low.}
 which would be directly produced via $e^+e^-\to \gamma a$. As in the case of the $A'$, also here the production and the decay are controlled by the same coupling, leading to too small production rates if the particle is to be long-lived. The exception to this conclusion are decays to the dark sector, followed by subsequent decays back to the Standard Model, a possibility that we now discuss in more detail.

\subsection{Emerging jets}\label{sec:emjets}
If a dark sector contains several relatively light states, the production of dark sector particles can result in dark jets with some of the dark sector particles decaying back to the Standard Model.\footnote{Such decays are possible even if the interactions with the Standard Model are very weak, \emph{i.e.}, even if the dark sector is in a ``hidden valley'' \cite{Strassler:2006im}.} A well motivated example is a dark sector that contains ``dark quarks'' $q_D$ charged under a confining force $SU(N_D)$. In this case the production of dark sector particles mimics the production of QCD jets in the Standard Model. The hard production process, $e^+e^-\to q_D \bar q_D$, is followed by the emission of ``dark gluons'' splitting into other dark quarks. This decay chain results in jets of dark particles, just as $e^+ e^-\to q \bar q$ production results in two QCD jets. The dark gluons and dark quarks confine into dark mesons, $\pi_D, \rho_D, \ldots$, and dark baryons, $n_D, n_D^*, \ldots$ The dark mesons decay to SM states, because the dark quarks have feeble interactions with the Standard Model. The lightest dark baryon is assumed to be stable and escapes the detector. 

The phenomenology of dark jets depends on the details of the dark sector model, both on the confining group, as well as on the dark sector field content and their masses (see Ref.~\cite{Knapen:2021eip} for some benchmarks). The pattern of decays into SM particles depends on the flavor structure of the portal interactions.  In principle many decays are possible, such as $\pi_D\to e^+e^-,\, \mu^+\mu^-,\, q\bar q,\, \gamma\gamma,\, \ldots$ and $n_D\to e^+e^- +{\rm inv},\, \ldots$ are possible. The branching ratios of the different channels are fixed by the flavor structure of the couplings to the SM fields. Examples of such signatures are {\em lepton jets} \cite{ArkaniHamed:2008qp,Baumgart:2009tn,Falkowski:2010cm}, {\em semi-visible  jets} \cite{Strassler:2008fv,Cohen:2015toa,Cohen:2017pzm}, {\em jet substructure from dark sector showers} \cite{Cohen:2020afv},  and {\em emerging jets} \cite{Schwaller:2015gea}. 

If the decays of dark sector particles are prompt, this results in a shower of visible particles in the \belletwo detector. The possibility that the decays lead to displaced vertices is experimentally less constrained. In this case the dark jets originate from many dark particles with varying decay times, resulting in a collection of displaced decays that appear in the detector as emerging jets forming far away from the interaction point~\cite{Schwaller:2015gea}. For large enough decay times the decays occur mainly outside the \belletwo detector, leaving only a signature of missing energy for \belletwo. A far detector like \gazelle would increase the volume in which the emerging decays can be detected, thus covering a larger range of possible decay times. A detailed study of the \belletwo sensitivity to such models can be found in \cite{Bernreuther:2022jlj}.

 As an example let us consider a simple toy model of cascade decays in the dark sector. A  dark scalar $\phi_1$
 is produced in positron--electron collisions, $e^+ e^-\to \phi_1 \phi_1$. Subsequently, $\phi_1$ decays to a pair of lighter dark scalars, $\phi_1\to \phi_2 \phi_2$, followed by $\phi_2$ decaying to SM particles, $\phi_2\to\,$SM+SM. The production of the final SM states, \emph{i.e.}, the constituents of emerging jets, is no longer controlled by a simple exponential as in the benchmark models in Section~\ref{sec:models-intro}, but rather by a convolution of two sequential decays. In the long lifetime limit the average probability for emerging jet constituents to arise in the detector is parametrically given by (rather than Eq.~\eqref{eq:P:parametric})
\begin{align}
     \langle\mathds{P}\rangle \approx \Omega \times \frac{D}{d_1}\times \frac{D}{d_2}\,,
\end{align}
with the respective decay lengths $d_{i}=\gamma_i \beta_i c \tau_i$ and assuming that $\phi_{1,2}$ are much lighter than the collision energy. Because of the larger radial size \gazelle has a parametric advantage to explore such scenarios compared to \belletwo. One can thus easily imagine a possibility where dark sectors are discovered in a missing energy signature at \belletwo and then explored fully only using \gazelle or a variation thereof.

\subsection{Quirks}\label{sec:quirks}
If the confining $SU(N_D)$ dark sector only contains heavy dark quarks, $Q_D$, in the GeV range, at \belletwo the production via $e^+e^-\to Q_D \bar Q_D$ does not leave enough energy to create another $Q_D \bar Q_D$ pair from the vacuum. Unlike in the case of light dark quarks, Section~\ref{sec:emjets}, now the dark mesons cannot form. Instead, the heavy quarks are connected by a flux tube of dark gluons that act as a string connecting at the ends to the two {\em quirks}, $Q_D$ and  $\bar Q_D$ \cite{Kribs:2009fy,Cai:2008au,Kang:2008ea}. The quirks fly apart until the energy stored in the string tension is increased enough to stop the quirks. At this point the string pulls the quirks back toward each other. Equating the kinetic energy $E_{\rm kin}\sim \sqrt{s} = \mathcal{O}(10\,\text{GeV})$ of the $Q_D\bar Q_D$ pair and the potential energy associated with the $SU(N_D)$ confining scale $\Lambda_{\rm IR}$ gives an estimate of the typical length scale of the string~\cite{Kang:2008ea}, 
\beq
\ell \sim \frac{E_{\rm kin}}{\Lambda_{\rm IR}^2}\sim 10\,\text{m}\,\frac{E_{\rm kin}}{10\,{\rm GeV}} \Big(\frac{10 \,{\rm eV}}{\Lambda_{\rm IR}}\Big)^2.
\eeq
 In the $e^+e^-$ center-of-energy frame the two quirks appear as though they were oscillating back and forth, connected by a string. In the lab frame, the two quirks have an overall boost and therefore move through the detector, oscillating back and forth, and finally exit. During the oscillations the string and the quirks slowly shed energy by emitting SM particles, which can potentially be observed in the experiment. They also lose energy through interactions with the detector material and thus can even get stopped in the detectors \cite{Evans:2018jmd}.
 A far detector like \gazelle would probe flux tubes with longer string length $\ell$ than \belletwo alone. For $E_{\rm kin}\sim 10\,\text{GeV}$, this would mean that one would probe dark force confinement scales in the range $\Lambda_{\rm IR} \lesssim 10\,\text{eV}$.
 
\subsection{Soft bombs}\label{sec:soft:bombs}
Strongly coupled hidden valley models \cite{Strassler:2006im} may result in spherically symmetric distributions of soft particles with very high multiplicity, the so-called {\em soft bombs}~\cite{Knapen:2016hky,Harnik:2008ax}. In the perturbative regime of a non-abelian gauge theory such as QCD, the radiation in the evolution of the jet is either collinear or soft, since these are the phase-space regions with logarithmically enhanced parton splitting. However, when the 't Hooft coupling of the gauge theory is large, parton emission with a large momentum fraction can occur also at larger splitting angles. This results in a more isotropic phase-space distribution of dark particles. Once the dark sector states decay to the SM states, one obtains the typical event topology of a soft bomb: large numbers of spherically distributed soft SM particles with typical momenta as low as 10 to 100 MeV. If the decays to the SM particles are suppressed by a small coupling, they could be displaced and occur outside the \belletwo detector. The wave of soft particles, a.k.a. the ``belt of fire'', could then only be seen by the \gazelle detector.

\section{Summary}\label{sec:summary}
\noindent Based on our investigations of the physics potential of \gazelle, we draw the following general conclusions.

In models with feebly coupled long-lived particles, such as heavy neutral leptons, axion-like particles or dark Higgs bosons, we have shown numerically that \gazelle provides at most a modest gain in sensitivity over \belletwo. For long lifetimes (and thus small couplings), the expected reach at \gazelle is up to a factor of ${\mathcal O}(1)$ higher than at \belletwo,  assuming no backgrounds. We find that the \belletwo detector itself is already very sensitive to even tiny couplings, \emph{i.\,e.}, long lifetimes, of particles in the $100$ MeV to GeV region. Compared to \belletwo, the three realistic detector geometries for \gazelle we considered have smaller angular coverage, but (in two out of three) a larger radial depth. This results in comparable effective fiducial acceptances and in a comparable reach of \belletwo and \gazelle for LLPs with long lifetimes. For shorter lifetimes the LLPs mostly  decay inside the \belletwo detector and do not reach \gazelle in appreciable numbers. 

The high angular coverage, smaller boosts, and relatively small backgrounds distinguish \belletwo from other experiments where far detectors can provide a substantial sensitivity gain in the LLP searches. For experiments with a low angular acceptance, such as LHCb, a far detector with larger fiducial acceptance and lower background, such as the CODEX-b proposal, is clearly a valuable addition to probe LLPs with long lifetimes. At ATLAS and CMS the angular coverage is already high. Thanks to a much larger size, the MATHUSLA proposal has, despite smaller angular coverage, a similar acceptance for LLP decays in the long-lifetime limit as ATLAS or CMS. Nevertheless, MATHUSLA  has a significantly increased projected reach to LLPs with lifetimes above 10\,m since it would  operate in the low-background regime without trigger limitations. The much smaller FASER experiment placed far downstream from the ATLAS interaction point will be able to very efficiently search for light new physics produced in the forward direction in an essentially zero-background environment, improving the reach of ATLAS and CMS by orders of magnitude.
Such large sensitivity gains are not expected at \gazelle compared to \belletwo, because the latter already operates in a relatively low background environment. Due to smaller boosts, both \belletwo and \gazelle are expected to be sensitive to shorter lifetimes than the proposed far detectors at the LHC, thereby covering the apparent gap left by the combination of near and far detectors at the LHC \cite{Filimonova:2019tuy,Lanfranchi:2020crw}.

We highlight two exceptions where the addition of a far detector at \belletwo could yield substantial sensitivity gains.
 First, light LLPs directly produced in electron-positron collisions are emitted mainly along the beam line, a region of phase space that is not covered by the \belletwo detector. In this case a far detector in the forward region could fill the acceptance gap of \belletwo and enhance the sensitivity. Second, many models with a confining force in a hidden sector predict phenomena like emerging jets, flux tubes or high-multiplicity final states, which emerge at a distance from the production point and require a large detection area, changing the effective acceptance in favor of \gazelle. Searching for signs of hidden confinement could thus motivate the construction of a far detector, even if the near detector has good angular coverage.

Besides the fiducial acceptance, the relative sensitivity of \belletwo and \gazelle
 depends crucially on the background suppression that can be achieved in different channels. For \gazelle, our studies of backgrounds from neutral kaon decays and cosmic muons lead us to believe that they can be rejected to a low level by exploiting the direction and timing of the particle trajectories.  \belletwo, on the other hand, has a high potential to go beyond
 displaced vertex searches  and detect particles with very long lifetimes through signatures with missing energy. There is thus complementarity of
 LLP searches via displaced vertices at \belletwo and \gazelle, and via missing energy signatures at \belletwo, which merits more detailed studies within concrete 
 models.

Beyond \belletwo, our findings can guide the design of far detectors at future electron-positron experiments, the ILC (see \cite{Schafer:2022shi}), CEPC and FCC-ee (for the latter see the HADES proposal~\cite{Chrzaszcz:2020emg}). Since these are expected to operate similarly in a relatively low background environment, the potential far detectors should aim at high fiducial coverages (large angular coverage and radial depth). Only in this way it is possible to appreciably increase the sensitivity reach of far detectors to LLPs over what should already be possible to achieve with general purpose detectors near the interaction point. 

In conclusion, the construction of a far detector at \belletwo that would require only a modest disruption of the present infrastructure could improve the reach of LLP searches. However, in general the gain in sensitivity is not very large. 
This highlights the great potential that the searches for LLPs at \belletwo already have. In the event of a discovery, the construction of \gazelle would become highly motivated. It
 would 
 facilitate further studies of LLP decays, especially if these involve
 non-minimal or strongly coupled dark sectors. Finally, with significant civil engineering efforts a much larger angular coverage of \gazelle than considered here could be possible. In this
 case gains in sensitivity to LLPs of up to two orders of magnitude compared to \belletwo could be possible. \\[2mm]

{\bf Acknowledgements}
We thank Jared Evans for contributing to the initial stages of this study, and Vladimir Gligorov, Phil Ilten, Simon Knapen, and Dean Robinson for clarifications about CODEX-b sensitivities. We thank Finn Tillinger for noticing a mistake in our original calculation of the detector extensions, which lead to the corrections in this revised version. JZ acknowledges support in part by the DOE grant de-sc0011784. The research of SW is supported by the German Research Foundation (DFG) under grant no. 396021762–TRR 257.
RS acknowledges support of the DFG through the research training group Particle physics beyond the Standard Model (GRK 1940). The research of CGC, KSH, SD, SL, and TF is supported by the DFG through Germany's Excellence Strategy -- EXC 2121 ``Quantum Universe'' -- 390833306 and in part by the Helmholtz (HGF) Young Investigators Group grant no.\ VH-NG-1303.
CGC is supported by the Alexander von Humboldt Foundation. CH and SL are supported by the Natural Sciences and Engineering Research Council of Canada. MT acknowledges the financial support from the Slovenian Research Agency (research core funding No. P1-0035).

\begin{appendix}
\section{HNL production and decay}\label{app:HNL}
\noindent Here we report the relevant $N$ production and decay mechanism used in Sec.~\ref{sec:HNL}. A complete review of these mechanisms can be found in Ref.~\cite{Bondarenko:2018ptm}.

In our analysis we take the case of $\tau$-flavored HNLs, that is $U_\tau \neq 0$, $U_e = U_\mu = 0$ and $m_N < m_\tau$, which makes $\tau$ decays the dominant production channels. We consider the decays $\tau^\pm\to N\pi^\pm$, $\tau^\pm\to N\ell_\alpha^\pm\bar\nu_\alpha$ and $\tau^\pm\to N\pi^\pm\pi^0$. The latter is greatly enhanced by the vector $\rho^\pm$ resonance and its decay width can be well approximated by the two body decay $\tau^\pm\to N\rho^\pm$ (see Ref.~\cite{Bondarenko:2018ptm}). The widths are
\begin{align}
    \Gamma(\tau^\pm\to N\pi^\pm) &= \frac{G_F^2 f_\pi^2 m_\tau^3}{16\pi}|V_{ud}|^2 |U_\tau|^2 \left[ \left( 1 - y_N^2 \right)^2 - y_\pi^2 \left( 1 + y_N^2 \right)^2 \right] \sqrt{\lambda(1,y_N^2,y_\pi^2)}\,,\\
    \Gamma(\tau^\pm\to N\ell_\alpha^\pm\bar\nu_\alpha) &\simeq \frac{G_F^2 m_\tau^5}{192\pi^3} |U_\tau|^2 \left[ 1 - 8 y_N^2 + 8 y_N^6 - y_N^8 - 12 y_N^4 \log (y_N^2) \right]\,, \\
    \Gamma(\tau^\pm\to N\rho^\pm) &= \frac{G_F^2 g_\rho^2 m_\tau^3}{16\pi m_\rho^2}|V_{ud}|^2 |U_\tau|^2 \left[ \left( 1 - y_N^2 \right)^2 + y_\rho^2 \left( 1 + y_N^2 - 2 y_\rho^2 \right)^2 \right] \sqrt{\lambda(1,y_N^2,y_\rho^2)}\,,
\end{align}
where $G_F = 1.16\times10^{-5}~{\rm GeV}^{-2}$ is the Fermi constant and $|V_{ud}| = 0.974$ is the CKM matrix element of light quark mixing. The meson masses and decay constants are $f_\pi = 130.2$ MeV, $m_\pi = 139$ MeV and $g_\rho = 0.162~{\rm GeV}^2$, $m_\rho = 775.5$ MeV respectively. Here we defined $y_i = m_i/m_\tau$, while the function $\lambda$ is defined as
\beq 
\lambda(a,b,c) = a^2 + b^2 + c^2 - 2 a b - 2 a c - 2 b c\,.
\eeq 
In the second equation we took the limit $y_\ell\to0$ to simplify the expression.

Once produced, the HNL can decay via $Z$-mediated processes only into leptons or pions. Final states are visible in \gazelle only when charged particles are present.
This is the case for some of the considered final states, while other ones remain invisible.

We consider the invisible decays $N\to\nu_\tau\bar\nu_\beta\nu_\beta$ ($\beta = e,\mu,\tau$) and $N\to\nu_\tau\pi^0$ with widths
\begin{align}
    \Gamma(N\to\nu_\tau\bar\nu_\beta\nu_\beta) &= (1 + \delta_{\tau\beta} ) \frac{G_F^2 m_N^5}{768\pi^3} |U_\tau|^2\,,\\
    \Gamma(N\to\nu_\tau\pi^0) &= \frac{G_F^2 f_\pi^2 m_N^3}{32\pi} |U_\tau|^2 \left( 1 - x_\pi^2 \right)^2\,,
\end{align}
where $x_i = m_i/m_N$.

The visible decays we take into account are $N\to\nu_\tau\bar\ell\ell$ and $N\to\nu_\tau\pi^+\pi^-$. Similarly to the production case, the two pions channel is enhanced by the $\rho^0$ resonance and its decay width can be well approximated by the two body decay $N\to\nu_\tau\rho^0$. The visible widths are then
\begin{align}
    \Gamma(N\to\nu_\tau\bar\ell\ell) &\simeq \frac{G_F^2 m_N^5}{192\pi^3}|U_\tau|^2 \left( \frac{1}{4} - s_w^2 + 2 s_w^4 \right) \left( 1 - x_\ell^2 \right) \,, \\
    \Gamma(N\to\nu_\tau\rho^0) &= \frac{G_F^2 \kappa_\rho^2 g_\rho^2 m_N^3}{32\pi m_\rho^2} |U_\tau|^2 (1 + 2 x_\rho^2) (1 - x_\rho^2)^2\,,
\end{align}
where $s_w$ is the sine of the weak mixing angle and $\kappa_\rho = 1 - 2 s_w^2$. In the first equation we neglected terms of order ${\cal O}(x_\ell^4)$, which is a good approximation for the three benchmarks considered in Section~\ref{sec:HNL}.
\end{appendix}

\bibliographystyle{JHEP_improved}
\bibliography{main.bib}

\end{document}